\documentclass[
aps,prl,
amsmath,amssymb]{revtex4-1}

\usepackage[pdftex]{graphicx}
\usepackage{amsmath,amssymb,graphicx,epsfig,wrapfig,color}
\usepackage[T1]{fontenc}

\usepackage{epsfig}
\usepackage{subfigure}

\begin{document}

\title{Two-color spectroscopy of UV excited ssDNA complex with a single-wall nanotube photoluminescence
probe: Fast relaxation by nucleobase autoionization mechanism}

\author{Tetyana Ignatova}\affiliation{Physics Department, Lehigh
University, 16 Memorial Dr.~E., Bethlehem, PA 18015}

 \author{Alexander Balaeff}\affiliation{NanoScience Technology Center, University of Central Florida,
12424 Research Parkway, Suite 400, Orlando, FL 32826}

 \author{Ming Zheng}\affiliation{National Institute of Standards and Technology, 100 Bureau Drive,
Gaithersburg, MD 20899}

 \author{Michael Blades}\affiliation{Physics Department, Lehigh
University, 16 Memorial Dr.~E., Bethlehem, PA 18015}

\author{Peter Stoeckl}\affiliation{Department of Physics \& Astronomy, University of Rochester, 206 Bausch \& Lomb Hall, 
 Rochester, NY 14627
}


 \author{Slava V.~Rotkin}
 \affiliation{Physics Department,
Center for Advanced Materials \& Nanotechnology and Center for Photonics \& Nanoelectronics,
Lehigh University, 16 Memorial Dr.~E., Bethlehem, PA 18015 }\email{rotkin@lehigh.edu}

\begin{figure*}[h]
	\centering
	\includegraphics{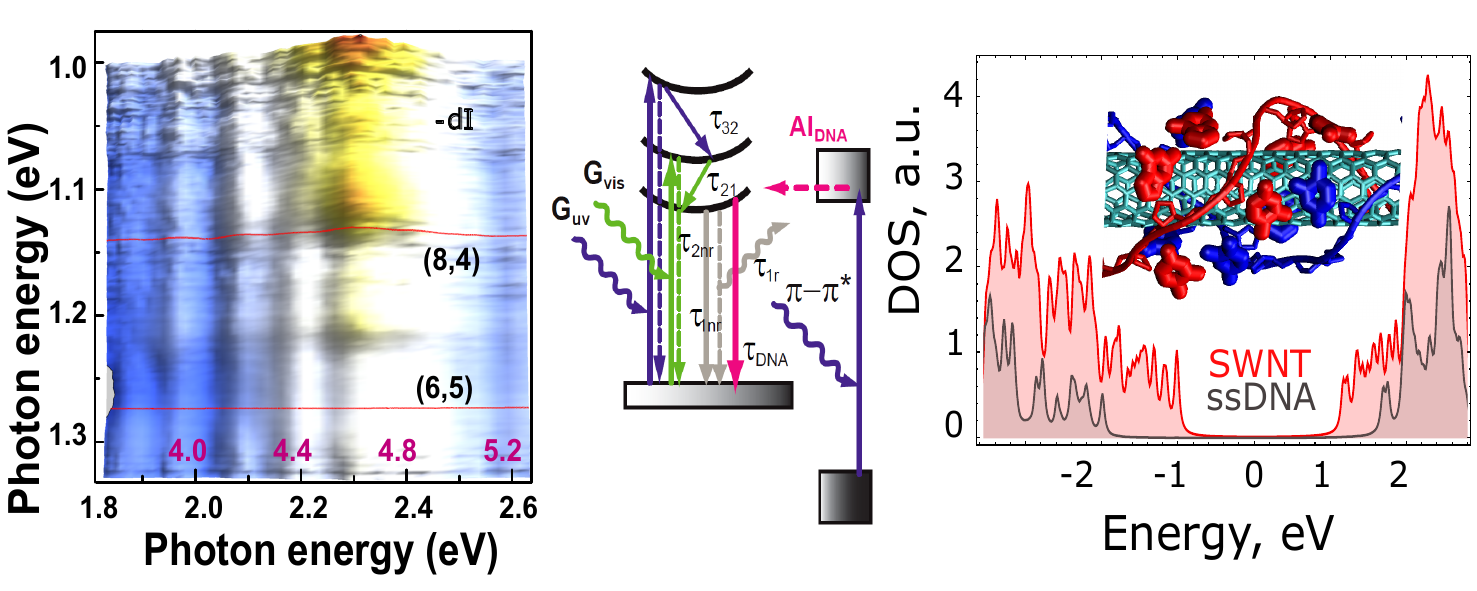}
\end{figure*}
DNA is capable to recover after absorbing ultraviolet (UV) radiation, for example, by autoionization (AI). Single-wall carbon nanotube (SWNT) two-color photoluminescence spectroscopy was combined with quantum mechanical calculations to explain the AI in self-assembled complexes of DNA wrapped around SWNT.

\vskip 2cm

\begin{abstract}
{
DNA autoionization is a fundamental process wherein UV-photoexcited
nucleobases dissipate energy by charge transfer to the environment without undergoing
chemical damage. Here, single-wall carbon nanotubes (SWNT) are explored as a
photoluminescent reporter for studying the mechanism
and rates of DNA autoionization. Two-color photoluminescence spectroscopy allows
separate photoexcitation of the DNA and the SWNTs in the UV and visible
range, respectively. A strong SWNT photoluminescence quenching is
observed when the UV pump is resonant with the DNA absorption, consistent with charge transfer from the
excited states of the DNA to the SWNT. Semiempirical
calculations of the DNA-SWNT electronic structure, combined with a
Green's function theory for charge transfer, show a 20 fs autoionization
rate, dominated by the hole transfer. Rate-equation analysis of the spectroscopy
data confirms
that the quenching rate is limited by the thermalization of the free
charge carriers transferred to the nanotube reservoir. The developed approach has a
great potential for
monitoring DNA excitation, autoionization, and chemical damage both {\it in vivo}
and {\it in vitro}
}
\end{abstract}

\maketitle





\section{1. Introduction}
\label{sec:intro}

It is well known that ultraviolet (UV) radiation presents a
significant danger for living organisms. DNA, which stores genetic information
in the majority of organisms on Earth, readily absorbs UV light. It has a signature absorption band at $\sim4.5-5$~eV ($\sim240-285$~nm)
resulting from the $\pi$-$\pi^*$ excitation of the nucleobases\cite{ARPC_2009}.
The amount of energy delivered by a single UV photon
(5~eV, corresponding to $\sim200$ 
times the thermal energy) is sufficient to trigger chemical reactions that corrupt
the DNA structure~\cite{DNA-radicals-Adhikary2005}. The cells with damaged DNA
either die or, even more
dangerously for a multi-cellular organism, begin uncontrollable (cancerous)
growth. To combat the damage, cells have developed DNA repair
mechanisms that, however, would be insufficient against a large UV radiation uptake
in the absence of alternative dissipation pathways\cite{ARPC_2009,rev-DNA-excit}.
Thus, the energy dissipation of the DNA
photoexcited states is crucial for the survival of DNA-based life.

Understanding the physics of DNA photoexcitation decay is important for such diverse fields as medicine,
evolutionary biology, and space exploration. For biomedical purposes, one strives to
understand the survivable levels of UV radiation for different cell types and learn the
ways to mitigate the irradiation effects. From the evolutionary perspective, the
energy dissipation mechanisms were crucial during the primordial cell evolution when
UV radiation on the young Earth was orders of
magnitude more intense than today while the DNA repair mechanisms were
presumably non-existent. For continued exploration of far space where
mankind may be capable to reach in the near future, it is crucial to develop
strategies for cellular and organismal safety in the harsh radiation conditions.

Known mechanisms of energy dissipation from DNA include direct singlet and indirect
triplet recombination, formation of secondary electrons, electron transfer and
deprotonation, and DNA autoionization (AI), that is, a spontaneous
irreversible charge transfer from the excited nucleobase to the environment. While the recombination mechanisms
have been studied quite extensively\cite{ARPC_2009,rev-DNA-excit,rev-DNA-excit,fast-PI-CT,fast-DNA}, DNA autoionization\cite{ai-photo} has received less
attention, partially due to the fact that AI in solution 
involves charge transfer from DNA to the surrounding water solvent\cite{dna07,review-e-hydr-DNA,fast-in-water}. The irregular, dynamically
changing structure of the solvent makes quantitative modeling of experiments and
theoretical predictions of AI rates and mechanisms extremely complicated\cite{e-in-water,e-solv-DNA}.

In this work, we explore the AI of single-stranded DNA (ssDNA) wrapped around
single-wall carbon nanotubes (SWNT) which constitute an elementary nanomaterial with a
great promise for such biotechnology applications as nanotherapeutics, nanopharmacology\cite{dai-review2015,NT-biomed_Dai2009}, and
optical imaging.
SWNTs are inexpensive and robust in synthesis, chemically inert, mechanically stable, and biocompatible\cite{bio-rev-NT2013,NT-cancer-drugDai2011}.
The existence of strong near infrared (NIR) SWNT bands that lie within the water transparency window make nanotubes remarkably suitable for bioimaging.
Indeed, transient absorption microscopy, single- and two-photon fluorescence, Raman, and photo-thermal microscopy have been
demonstrated both in vitro and in vivo\cite{bio-rev-NT2013,NT-Raman-cells-Heller2012,RamanNT-bio2014,NT-IR-biosens2015}.
Due to the nanotube transverse size of a few nanometers (nm) and large aspect ratio up to $10^4-10^6$, controllable, in principle, by synthesis and/or post-processing,
several applications in biological sensing have been achieved through direct mechanical, optical, and electronic interactions with biopolymers and
cellular organelles of a similar size\cite{Collins2012,NTmed_rev_Martin2003,NT-detector-Burke2015,heller}.

In particular, DNA interaction with SWNTs has been an object of intensive
study\cite{zheng03,zheng-DNA-NT2004,DNA-NT-Kaxiras2005,NT-DNA-Heller2006,tu09,khripin09,ARPC-Rotkin2010,DNArecognition-Jagota2012}
Chemical engineering and bioengineering applications of  DNA-SWNT complexes have been suggested\cite{weisman,strano,NT-bioeng2012,NT-protein-rev2012,rev-NTbioeng-2014}.
ssDNA is known to wrap in a regular pattern around the cylindrical surface of
the SWNT\cite{zheng03,zheng-DNA-NT2004,DNArecognition-Jagota2012,NT-DNA-Zheng2003}. The nucleobases bind to the nanotube surface through a
combination of van der Waals, hydrophobic and Coulomb forces\cite{jagota,stacy,NT-DNA-RNA-stable-in-water,jagota-Roxbury2013}, while
the phosphodiester backbone remains exposed to the solvent. The wrapping pattern
depends strongly on the DNA sequence\cite{tu09,Schulten}.

\begin{figure}[htb]
	\centering
	{   \includegraphics[width=3.2 in]{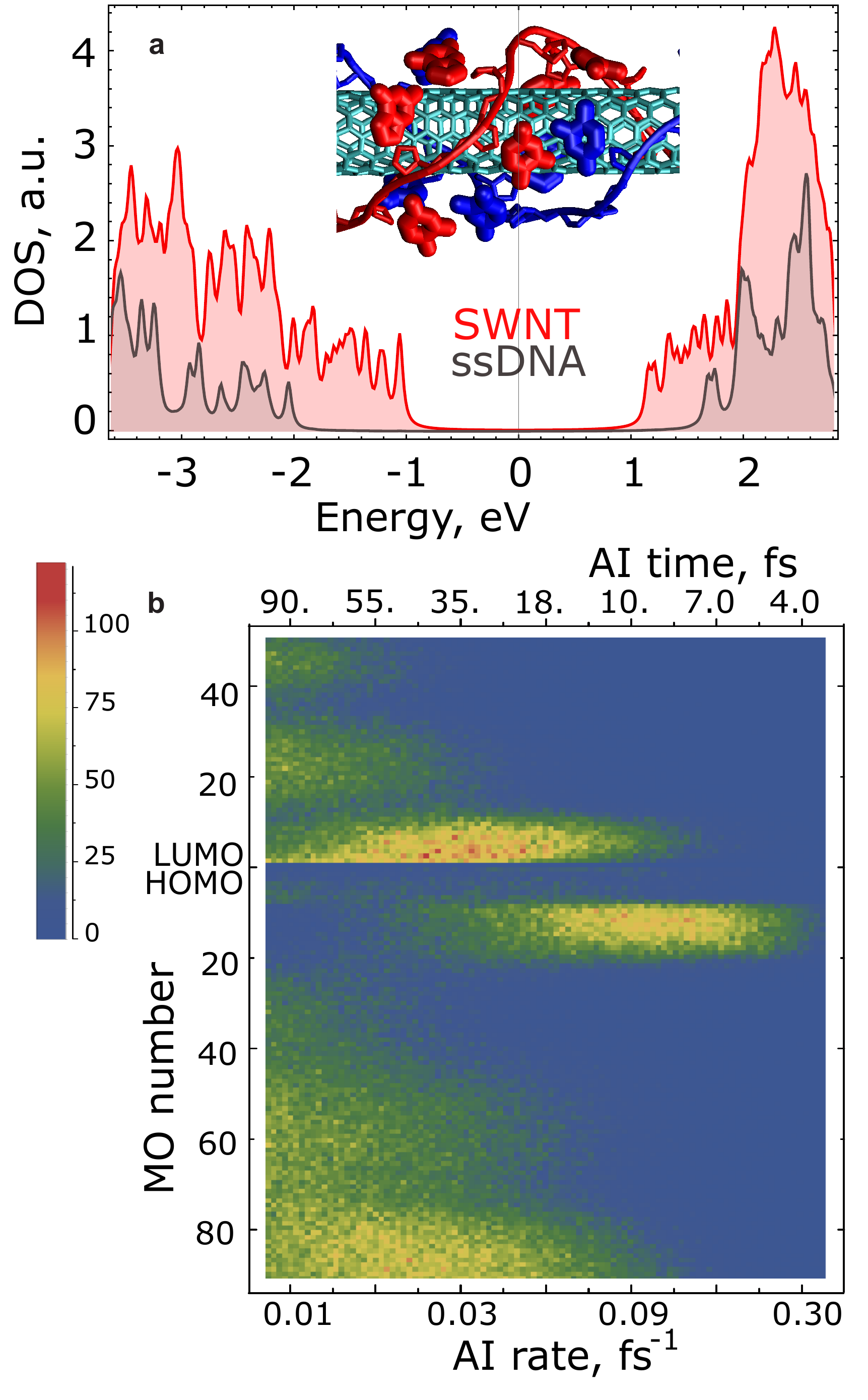}
		\caption{(a) CNDO partial density of states (DOS) of the DNA (gray) and SWNT
			(pink) near the charge neutrality point, calculated
			for one unit cell of a (6,5) nanotube wrapped with two intertwined
			strands of  poly(GT) DNA (red and blue in the inset).
			The DOS was calculated using an empirical line broadening of 30~meV and
			a gap rescaling factor of 1.67.
			(b) Statistical distribution of the AI rates for an electron excited in the
			50 lowest unoccupied MOs (LUMO) and for a hole in the 90
			highest occupied MOs (HOMO). Note the scales of the upper and the lower axes
			are the reciprocals of each other.
		}
		\label{fig:figA}}
\end{figure}

The geometry of the DNA-SWNT complex results in a strong electronic interaction
between the nucleobases and the nanotube. The ionized backbone of the helically
wrapped DNA may alter the nanotube electronic structure via symmetry
breaking~\cite{puller,spain,stacy-optics}. Modulation of the optical and electronic properties of
SWNTs has been experimentally observed upon DNA
wrapping\cite{zheng-DNA-NT2004,tu09,DNA-NT-Golovchenko2007,other-DNA-NT-PL2008,duque09,fagan13}.
Recently, SWNTs were found\cite{zheng-oxid} to mitigate the ssDNA oxidative damage
caused by the radical species in water ultrasonication experiments, at least in {\em in
vitro} conditions. The latter study stresses that the charge transfer between DNA and
nanotube is likely responsible for the protection effect.
The existence of such a charge transfer mechanism would be consistent with the well-known
photoinduced charge transfer both within DNA and between DNA and other species 
\cite{fast-PI-CT,ZHAN2014,GIESE01,BARN2001,SENT2005,LEWI1997,GENE2010,LEWI2006,RENA2013}.
A large overlap of SWNT conduction and valence bands with the DNA $\pi^*$ (and
$\pi$) bands (Fig.\ref{fig:figA}) 
should allow for both charge transfer and
photoinduced modulation in DNA-SWNT systems, which has not been studied before.

\begin{figure}[htb]
	\centering
	{   \includegraphics[width=3.2 in]{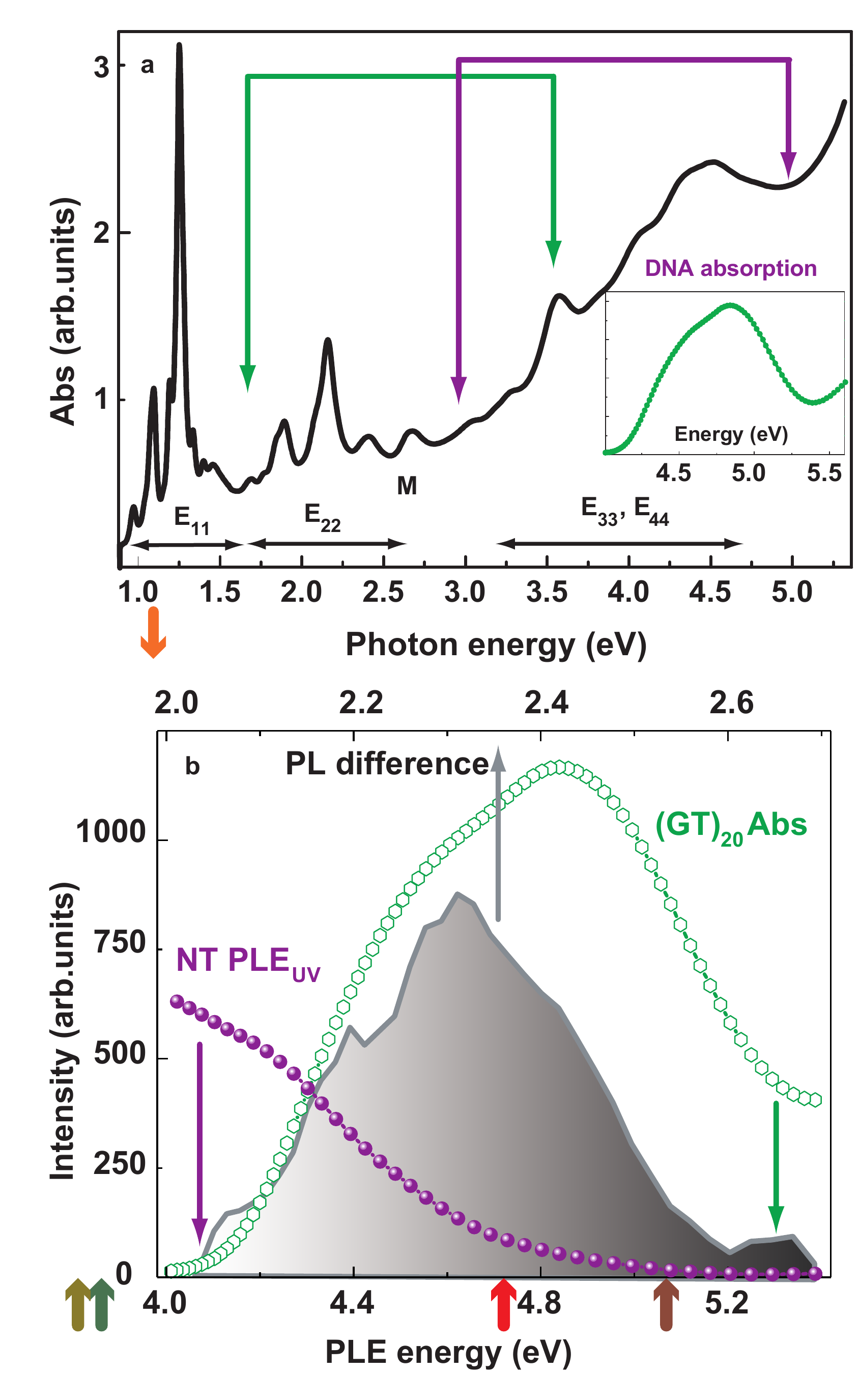}
		\caption{(a) Absorption spectrum of DNA-SWNT solution. The SWNT absorption
			subbands are labeled by $E_{nn}$. The green and purple arrows indicate
			the two-color PL excitation range. The inset shows the absorption
			spectrum of (GT)$_{20}$ ssDNA\cite{dna-abs-paper}. (b) The UV part of
			the PLE spectrum of the nanotube solution (purple solid symbols) taken
			at PL photon energy $E=1.11$~eV (indicated by the orange arrow in the top panel
			and Fig.\ref{fig:fig2}b), overlaid with the DNA absorption from the top panel
			(green open symbols) and $\delta I$ (gray line), measured at half of the photon
			energy (note the different scale of the top axis).
		}
		\label{fig:fig1}}
\end{figure}

Due to the high symmetry of the SWNT lattice, its electronic states
form bands with well defined quantum numbers (related to axial and angular
momentum)\cite{dresselhaus,ando1993}. The rigidity of $sp^2$-carbon bonds makes SWNTs
stable and their electronic structure mostly intact even
in solvent\cite{NT-DNA-RNA-stable-in-water}. As a result, the SWNT optical transitions
have been very well characterized experimentally, in contrast to the states
of electrons trapped in the water/solvent. Thus, the nanotube provides an ideal probe for studying the
nucleobase autoionization mechanism and consequently, the non-chemical energy dissipation
pathway for the UV photoexcited states of DNA. 
Electronic states of the SWNT can be monitored through the SWNT UV-NIR absorption (Fig.\ref{fig:fig1}) and
photoluminescence (PL) response  (Fig.\ref{fig:fig2}) in the NIR
range, completely outside of the DNA excitation (absorption) spectrum
(Fig.\ref{fig:fig1}, inset).

Here, we demonstrate that the PL response of the nanotube (as a model optical system) in the DNA-SWNT
complex is strongly modulated by the photoexcitation of the wrapped DNA. Two-color UV/visible excitation
spectroscopy is used to excite the CoMoCat nanotube sample wrapped with (GT)$_{20}$
ssDNA\cite{05_Ignatova2011}. SWNTs and the nucleobases were excited independently
using two harmonics of the diffraction grating (see 
 section \ref{sec:methods}).
 The UV-excited DNA is
shown to undergo an AI charge transfer to the SWNT on the time scale of 20 fs, which dominates
all other decay mechanisms for the DNA $\pi$-$\pi^*$ state. The transferred free charge
carriers, both electrons and holes, contribute substantially to the
non-radiative recombination of the SWNT
excitations, quenching its PL response. The rate equation analysis of the spectroscopy data
yields the PL quenching rate on the order of 40-60 ps, much slower than the AI rate, most likely limited
by the SWNT electron thermalization. The ability of living cells to uptake DNA-functionalized SWNTs\cite{NT-endocyt-Strano2008}
makes the two-color spectroscopy method suitable for monitoring DNA autoionization in vivo.

\section{2. Methods}
\subsection{2.1. Experimental Methods}
\label{sec:methods}

The samples were prepared using the procedure described in detail
elsewhere. In brief, SWNTs synthesized with the CoMoCat
method, were wrapped with (GT)$_{20}$ ssDNA\cite{05_Ignatova2011},
citric buffer was substituted from the solution, and the sample was re-suspended in
heavy water (D$_2$O) to remove any trace of the original buffer and free
DNA\cite{zheng03}. The absorption spectrum of the resulting system (Fig.\ref{fig:fig1}a) shows clear $E_{11}$,
$E_{22}$, $E_{33}$, and $E_{44}$ transitions for
individual SWNT species
. The peak centered at
265~nm (4.68~eV) corresponds to the signature DNA absorption due to
$\pi$-$\pi^*$ transitions. A theoretical absorption curve for (GT)$_{20}$ DNA, shown in
the inset, matches the peak well.

\begin{figure}[htb]
	\centering
	{   \includegraphics[width=3.2 in]{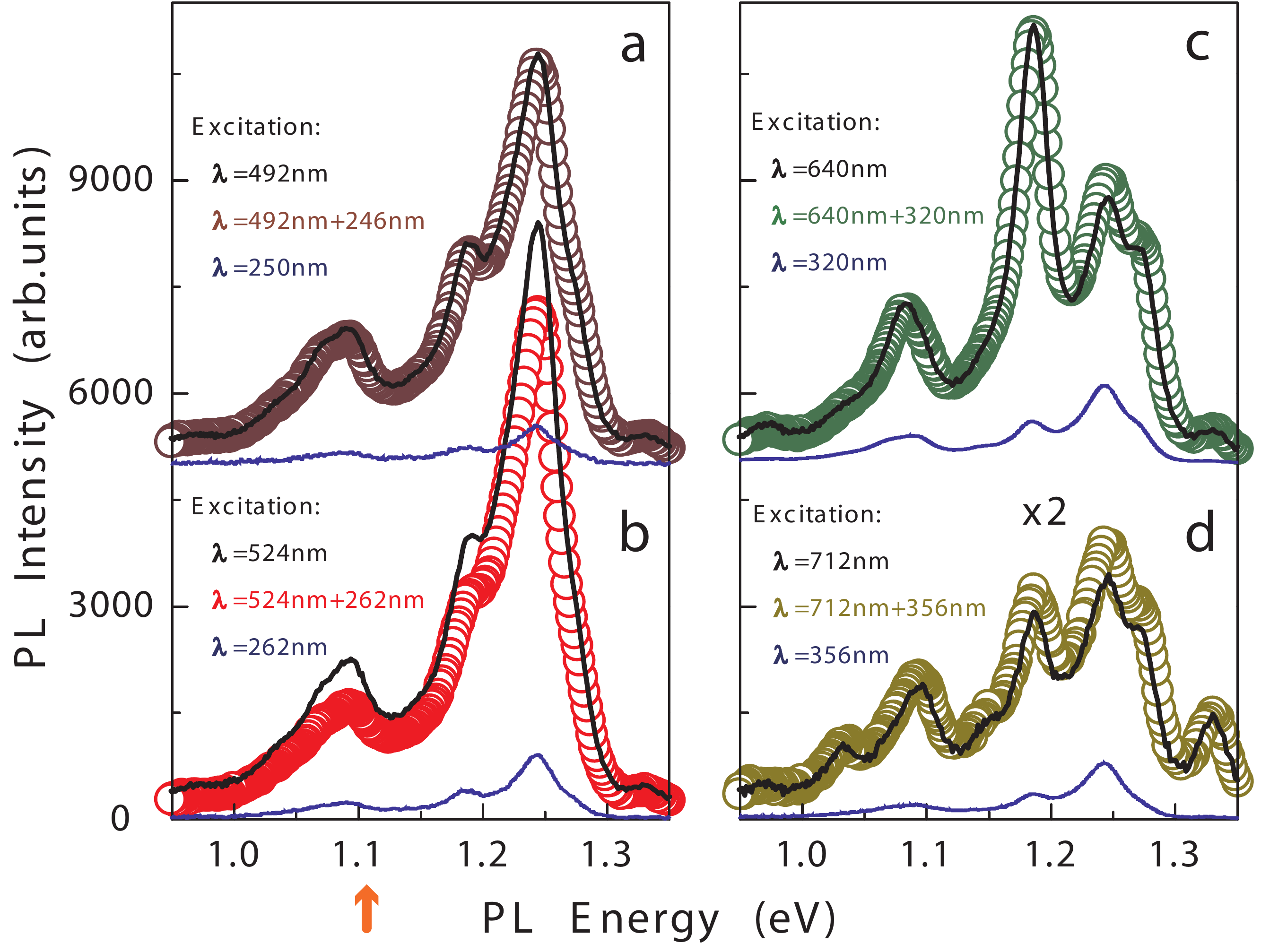}
		\caption{Single- and two-color PL spectra of DNA-SWNT solution. Legend shows
			excitation wavelengths: each panel presents PL emission induced by 1-color
			visible
			excitation (thick black line), 1-color UV (thin blue line), and 2-color combined UV and visible excitation
			(open symbols), taken at PLE energies indicated by the same color arrows in
			Fig.\ref{fig:fig1}. Orange arrow indicates PL energy where the UV PLE data was taken
			in Fig.\ref{fig:fig1}.}
		\label{fig:fig2} }
\end{figure}

A 2D photoluminescence/photoluminescence excitation (PL/PLE) plot for the
DNA-SWNT solution was taken by a FLUROLOG5 fluorimeter in a wide range of excitation
from 250 nm to 750 nm  (Fig.\ref{fig:fig4}d and SI). $E_{11}$ PL transitions
show a Stokes shift of about 5~nm with respect to the absorption peaks, typical for
solutions of DNA wrapped SWNTs\cite{Hertel2010}. All
experiments were performed in ambient at room temperature; the excitation
intensity was kept in the linear regime.

The 2$^{nd}$ harmonic of the diffraction grating turned to the appropriate
grazing angle was used for the second-color (UV) excitation, {\em i.e.}, the
wavelength of the UV source was half that of the main excitation
$\lambda_{\textrm{uv}} = \lambda_{\textrm{vis}}/2$. The intensity of the 2$^{nd}$ order (UV) constituted
approximately 10\% of the intensity of the first order (in the visible range) or
slightly less. Reference experiments were performed with only the visible range
excitation and the UV light completely blocked on the excitation pass by a UV long
pass filter (UVLP 450, Thorlabs). Measured transmission spectral response of the
UVLP filter (see SI) was used to correct the excitation intensity.

\begin{figure}[htb]
	\centering
	{   \includegraphics[width=3.2 in]{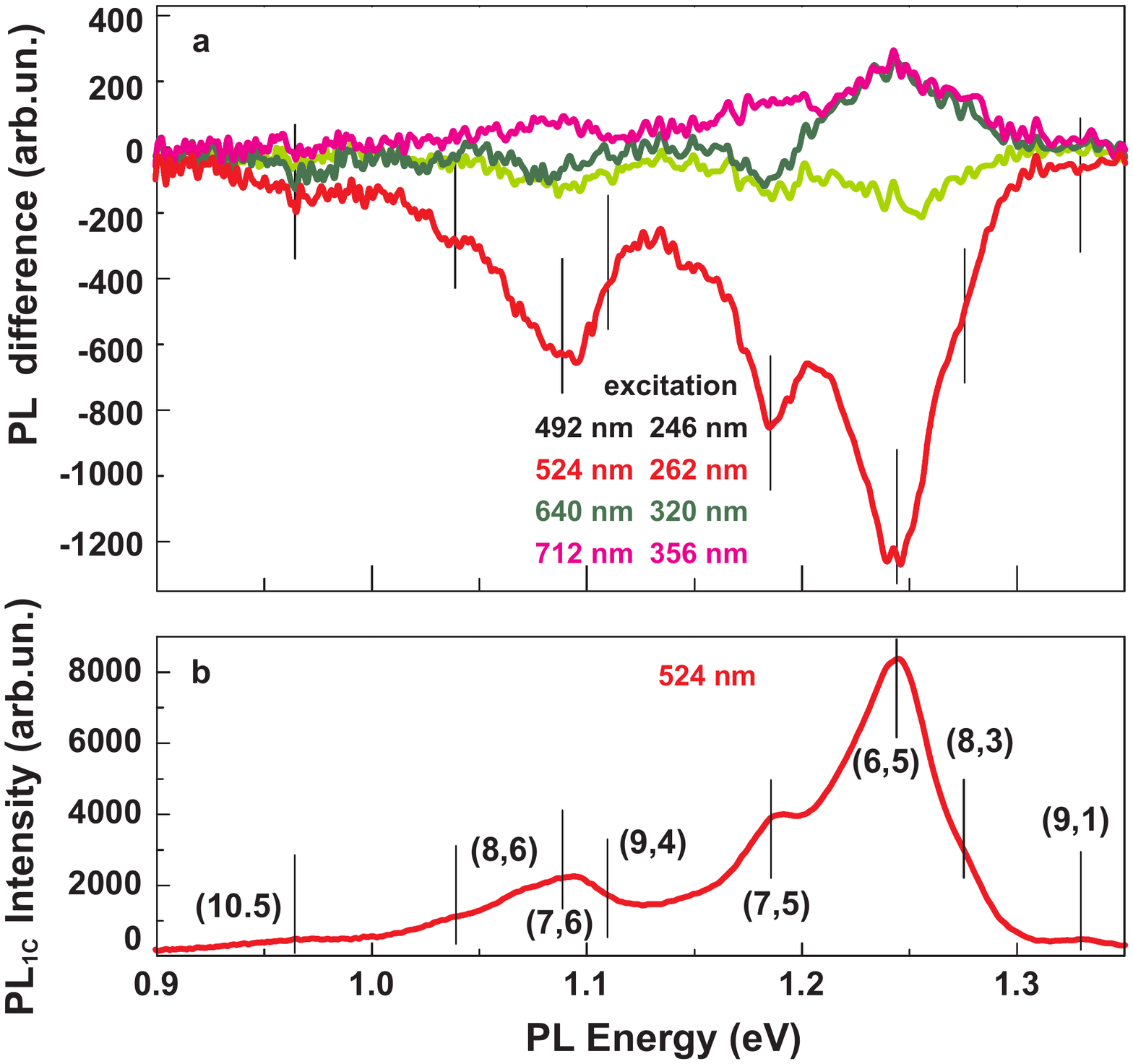}
		\caption{(a) PL difference ($\Delta I$) for the same two-color excitation lines as in Fig.\ref{fig:fig2}: 492/246 nm, 524/262 nm,
			640/320 nm, 712/356 nm (brown, red, gray, green). Negative differential PL at
			additional UV-excitation in the band around 4.6~eV (265 nm) is observed, corresponding to the DNA absorption band (Fig.\ref{fig:fig1}).
			(b) PL spectrum with one-color excitation at 2.36~eV (524 nm), showing the same spectral features as the differential PL in top panel; each line is
			identified with an individual SWNT chirality.
		} \label{fig:fig3} }
\end{figure}

\subsection{2.2. Theory and simulations}
\label{sec:theory}

The electronic interactions between the SWNT and the wrapped DNA were studied using
the semi-empirical quantum chemical modeling method
INDO/s~\cite{RIDLEY1973,VOITYUK2013}, as implemented in the
program CNDO\cite{CNDO2008}. The INDO/s method has been used in the past to calculate the
electronic properties of DNA, as well as those of carbon
nanotubes~\cite{ZHAN2014,VENK2011,VENK2011A,HATC2008,FONS1996,VOIT2007,VOIT2008},
and showed
accuracy comparable to that of higher-level quantum calculations\cite{VOITYUK2013,THIE2014,VOIT2006}.
The INDO/s electronic structure was calculated for each of the 6200 frames of the 3.1~ns molecular dynamics
(MD) simulation\cite{jagota-Roxbury2013} of two intertwined (GT)$_{30}$ DNA chains wrapped around a 70~\AA-long
fragment of (6,5) SWNT (see Fig.~\ref{fig:figA}a). Note that (6,5) is the most abundant SWNT species in
our CoMoCat samples. In view of the computational resource limitations, the CNDO
calculations were performed for only a part of the MD system that included a single
unit cell of the SWNT (41~\AA~in length) and the DNA bases overlaying the unit cell with their
coordinates extracted from the MD frames (see SI for detail).
The dangling bonds from the SWNT carbon atoms and the
nucleobase N atoms were capped with hydrogens\cite{TheKilins}.
The resulting quantum mechanical
system included 386 SWNT atoms (364 C atoms and 22 capping H atoms)
and, depending on the frame, 14-20 nucleobases (capped with H atoms).

Three CNDO calculations were performed for each MD frame: one for the complete
system (DNA and SWNT), one for only the SWNT section, and one for only the
nucleobases included. The complete system calculation results in ``hybridized'' molecular orbits (MOs), $|\Phi^{hyb}\rangle$,
delocalized between the DNA and the SWNT lattice. 
The DNA-only and
SWNT-only calculations result in ``unhybridized'' MOs, localized on either the DNA, $|\Phi^{DNA}\rangle$,
or the SWNT, $|\Phi^{NT}\rangle$. The ``unhybridized'' MOs, put together, result in a (generally, unorthogonal) orbital
basis set:
\begin{eqnarray}
\displaystyle |\Psi\rangle = \sum_{i=1}^{N_{DNA+NT}} a_i |\Phi^{hyb}_i\rangle = \nonumber\\
\sum_{i=1}^{N_{DNA}} b_i |\Phi^{DNA}_i\rangle + \sum_{i=1}^{N_{NT}} c_i |\Phi^{NT}_i\rangle,
   \label{hyb-basis-set}
\end{eqnarray}
that allows one to determine the density of electronic states on
the DNA and the SWNT:
\begin{eqnarray}
DOS(E,DNA) = \sum_{i}^{N_{DNA}} |b_i|^2 \delta(E-E_i), \\
DOS(E,NT) = \sum_{i}^{N_{NT}} |c_i|^2 \delta(E-E_i).
\label{mo_hyb_vs_unhyb}
\end{eqnarray}

\section{3. Results}
\label{sec:results}
\subsection{3.1. DNA autoionization rate: Theoretical consideration}

The electronic structure of the DNA-SWNT system was used to evaluate the
rate of charge transfer\cite{CT-Ratner2011} of an electron (or a hole) from a photoexcited nucleobase to
the SWNT. Such a process describes both the AI mechanism (if only one charge carrier
is transferred onto the SWNT) and the resonance energy transfer (if both the
electron and the hole move onto the nanotube).
Since the number of atoms (and therefore the number of delocalized electrons) of
SWNT exceeds by far that of a single DNA nucleobase, the nanotube plays the role of
the reservoir for the DNA charge carriers transferred onto it. Such charge carriers
quickly thermalize on the SWNT and become indistinguishable from other charge carriers in the nanotube.

 For each MD frame, we assume that the
system at time zero occupies the photoexcited state which comprises an electron in
one of the unoccupied MOs and the hole in one of the occupied MOs of the DNA-only
subsystem. The evolution of the photoexcited state is then calculated using the
Green's function technique 
(see the supplement for detail).

In brief, it can be shown that the probability of autoionization by time $t$ is the sum of the
probabilities of electron and hole transfer from the DNA to the SWNT:
\begin{equation}
{\cal P}={\cal P}_e+{\cal P}_h = G_e(t)G_e^+(t) + G_h(t)G_h^+(t)
   \label{prob-AI}
\end{equation}
where $G_e(t)$ and $G_h(t)$ are the Green's functions for the charge transfer from the DNA to the SWNT for electron and hole, respectively:
\begin{eqnarray}
G_e(t) = \sum_{j_1,j_2}\sum_{k_1,n_2}  \langle\Phi^{NT}_{k_1} | e^{-iHt/\hbar} |\Phi^{DNA}_{j_1}\rangle
\nonumber\\ \times
\langle\Phi^{DNA}_{j_2} | e^{iHt/\hbar} |\Phi^{DNA}_{n_2} \rangle
\label{GreensFn-e}
\end{eqnarray}
and
\begin{eqnarray}
G_h(t) = \sum_{j_1,j_2}\sum_{n_1,k_2}   \langle\Phi^{NT}_{k_2} | e^{-iHt/\hbar} |\Phi^{DNA}_{j_2}\rangle
\nonumber\\ \times
\langle\Phi^{DNA}_{j_1} | e^{iHt/\hbar} |\Phi^{DNA}_{n_1} \rangle.
\label{GreensFn-h}
\end{eqnarray}
Here, $H$ is the Hamiltonian of the DNA+SWNT system; the indices $j_1$ and $j_2$ run
over the unoccupied and occupied DNA orbitals, respectively, and correspond to the
electron and the hole placement in the initial excited state;
the indices $k_1$ and $k_2$ label the unoccupied and occupied orbitals of the SWNT and the indices $n_1$ and $n_2$
run over the unoccupied and occupied orbitals of the DNA, all in the final state. Such a formalism takes into account not only the
charge transfer of the electron (hole) from the DNA to the SWNT, but also the transfer of
the remaining hole (electron) between the DNA states\cite{CT-in-DNA} (see the supplement for further
detail).

The time-dependent probability of AI from specific DNA MOs, both occupied and
unoccupied, was computed along the MD trajectory. The
statistical analysis of the results is presented in Fig.\ref{fig:figA}b.
Each row of the density plot in the figure represents
a probability distribution function for the partial AI rate from a particular
HOMO+N, N$^{th}$ occupied MO (or LUMO+M, M$^{th}$
unoccupied MO) state of the DNA, where $1\le N\le 90$ and $1\le M\le50$.
A broad (order-of-magnitude) distribution of rates for each MO results from the MD
fluctuations of the DNA-SWNT complex geometry.
Most efficient AI is observed from two clusters of MOs closest to the HOMO-LUMO gap
(corresponding to individual photoexcited nucleobases). Besides this, statistical
analysis clearly shows that the AI rates for hole transfer are faster (about 5
fs) than those for electrons (15 fs). Next we estimated the total probability of the
UV-ionization of DNA by convolving the AI probability with the theoretical DNA
photoexcitation (absorption) spectrum. For a given wavelength
$\lambda_{\textrm{uv}}$ of the UV pump, the total AI probability, related to the
experimentally measured DNA charge transfer rate to be discussed below, is expressed
as:
\begin{equation}
AI(\lambda_{\textrm{uv}})=\sum_{j_1,j_2} {\cal K}^{(j_1,j_2)}(\lambda_{\textrm{uv}})\, {\cal P}_{j_1,j_2}(t),
 \label{spectral-ai-rate}
\end{equation}
here ${\cal P}_{j_1,j_2}$ are the partial AI rates from
Eq.(\ref{prob-AI}-\ref{GreensFn-h}) for a given electron-hole pair at MOs $(j_1,j_2)$, and
${\cal K}^{(j_1,j_2)}$ is the DNA photoexcitation rate between these MOs at $\lambda_{\textrm{uv}}$.

Fig.~\ref{fig:fig4}f summarizes our findings: the fast and efficient AI is
clearly seen in the spectral band near 265~nm (4.7~eV). 
It takes less than 20 fs for a
significant number of photoexcited charge carriers to tunnel into the SWNT. Our method allowed us to
explore whether the SWNT conduction bands accept the photoexcited electron from the DNA base
(e-tunneling) or the SWNT valence bands accept the hole from the DNA valence states
(h-tunneling). Fig.\ref{fig:figA} shows that hole
transport for the modeled DNA-SWNT complex is more efficient: the AI lifetimes
are 5-25 fs for the holes and 15-65 fs for the electrons.

A direct experimental observation of such fast AI dynamics would require a transient-state
characterization method with fs resolution. However, for the DNA-SWNT hybrid the AI
can be inferred from the steady-state population of the charge carriers
transferred to the SWNT, given that the mechanism is
not overwhelmed by other processes. The latter is a safe assumption since the DNA nucleobases are
strongly hybridized to the SWNT, so the average DNA coupling
to the solvent is much smaller than to the nanotube reservoir,
and the other non-radiative and radiative decay channels are known to have
even longer characteristic times\cite{ARPC_2009}.
Thus, the AI-driven depopulation rate of the photoexcited state of the DNA equals
the population rate of the SWNT. This charge transfer is followed by the
thermalization of SWNT hot free carriers and then PL quenching. The latter can be directly extracted from the
two-color PL experiments, as explained next.

\subsection{3.2. Nanotube as a two-color photoluminescence probe}

A sketch of the band structure of the DNA-SWNT
hybrid is shown in Fig.\ref{fig:fig4}a. In normal conditions, resonant excitation, $G_{\textrm{vis}}$,
creates a nanotube exciton in the $E_{22}$ manifold. The exciton undergoes quick
relaxation to the first subband followed by a nonradiative recombination
with the rate $\gamma_{nr1}$ or a radiative recombination with the rate
$\gamma_{r1}$, giving rise to PL with a quantum yield (QY) proportional to the ratio $\sim\gamma_{r1}/\gamma_{nr1}$.

In our two-color excitation scheme, standard visible excitation is combined with an additional UV pump,
$\tilde{G}_{\textrm{vis}} + \tilde{G}_{\textrm{uv}}$. The latter can be
tuned to the resonant excitation of either ssDNA or nanotube, while the former excites only SWNT species in the complexes.
The nanotube resonant UV excitation generates additional e-h
pairs in the $E_{33}$ subband that can relax to $E_{22}$ and subsequently to
$E_{11}$ subbands, and recombine radiatively from the lowest energy
level, contributing to additional PL (Fig.\ref{fig:fig4}a).

Evolution of the SWNT PL upon adding a small UV pump
at different wavelengths of excitation shows three qualitatively
different types of behavior (Fig.\ref{fig:fig2}). While additional UV illumination
causes (\emph{i}) a higher PL intensity within a wide range of $\lambda_{\textrm{uv}}$
(Fig.\ref{fig:fig2}c-d), in a certain range the PL intensity has (\emph{ii}) nearly the same
value with and without second color excitation (Fig.\ref{fig:fig2}a). This correlates with the UV pump
not being in resonance with one of $E_{33/44}$ SWNT transitions. Thick black
solid curves in Fig.\ref{fig:fig2} show SWNT emission spectra taken with the
photoexcitation only in visible, at $\lambda_{\textrm{vis}}$ (UV pump was
completely blocked). The (green/red/brown) open symbols correspond to
dual-color excitation (by both visible and UV source),
$\tilde{G}_{\textrm{vis}} + \tilde{G}_{\textrm{uv}}$. Thin curves, shown
at the bottom of each plot, correspond to the
SWNT emission observed at pure UV single-color excitation,
$\lambda_{\textrm{uv}}$, scaled according to UV source efficiency. We emphasize that in the first
two cases (\emph{i}-\emph{ii}) the dual-color-excitation PL (green and
brown symbols) can be found as a simple sum of visible excitation spectra
(thick lines, without UV pump) and UV excitation (thin, without visible pump): $PL(\tilde{G}_{\textrm{vis}}) + PL(\tilde{G}_{\textrm{uv}})$.

The PL behavior was found to be drastically different in a narrow band of
dual-color-excitation $\lambda_{\textrm{vis}} = 520-550$~nm/$\lambda_{\textrm{uv}} = 260-275$~nm. In this spectral
band the SWNT emission intensity shows (\emph{iii}) an anomalous drop
under an additional UV illumination despite the fact that the UV excitation is in resonance with
the SWNT bands (compare red symbols and black curve
in Fig.\ref{fig:fig2}b).

Fig.\ref{fig:fig3} compares the one-color PL (bottom) with both normal and
abnormal two-color PL for different SWNT species. The difference in PL intensity with and without additional UV
pump: $\Delta
I=I^{\textrm{vis}+\textrm{uv}}-\tilde{I}^{\textrm{vis}}$ is plotted in the top panel vs. the PL energy.

At $640/320$~nm the SWNT are
resonantly excited by both G$_{\textrm{vis}}$ and G$_{\textrm{uv}}$
pump (type \emph{i}). UV generation of
charge carriers in the higher $E_{33}$ and $E_{44}$ SWNT subbands is followed by
their relaxation from those subbands to the lowest subband, and radiative
recombination. At this particular wavelength of the second-color excitation (320 nm)
the PL enhancement is expected for (8,3), (6,5), (7,5) and (7,6) SWNTs
(see SI for details) which is fully corroborated by experimental data.

At $492/244$~nm UV pump $G_{\textrm{uv}}$ is not resonant with SWNT bands,
as shown in Fig.\ref{fig:fig2}a (thin curve). The number of additional e-h pairs due to UV excitation is
negligible and the PL difference (brown curve in Fig.\ref{fig:fig3}) is close to zero (type \emph{ii}).

The picture is qualitatively different at 520/260 nm of excitation (red curve). $\Delta I$
is negative, which means that UV illumination not only fails to generate
new excitons but also initiates an efficient PL quenching. We stress on three major observations: (1)
PL quenching happens uniformly for all semiconducting tubes present in
solution and the process has a similar strength for each of the PL lines
(compare differential PL emission, $\Delta I$, and one-color excitation
spectrum, $I^{\textrm{vis}}$ in Fig.\ref{fig:fig3}). All single-color PL features assigned to
particular chiralities are present in the same proportion in two-color $\Delta I$.  At the
same time, (2) this quenching process varies with the photon energy of
excitation: it manifests itself in a narrow PLE spectral range which is (3) not
resonant with any of $E_{33}$ or $E_{44}$ transitions of SWNTs present in
the sample. Absence of chiral selectivity in $\Delta I$ along with observation of
a distinct PLE band and its coincidence with the spectral absorption band of
the DNA at 265 nm indicates that an additional nonradiative
decay channel appears due to resonantly excited DNA. We note that the
effect was not observed in similar SWNT samples without DNA, dissolved using different
surfactants (Na deoxycholate/Na dodecylsulfate), while replacement of D$_2$O with H$_2$O
preserves the effect. Our samples do not contain other species, which
allows us to attribute its sole origin to DNA unambiguously.

\subsection{3.3. Rate-equation analysis of the SWNT PL quenching}

The DNA induced PL decay rate can be obtained quantitatively
from the measured PL and PLE data by solving the system of rate equations for the population of SWNT excited
states. A steady-state solution of the rate equations is readily obtained (see SI for derivation
details). In brief, we solve a 4-level system for a nanotube, adding the DNA AI pathway
via a spectral quenching rate $\gamma_{DNA}$. In order to prove that the derived PL
quenching rate is indeed caused by the DNA AI, the spectral profile of $\gamma_{DNA}$ has been
calculated and compared to the DNA absorption (photoexcitation) spectrum.

The two-color excitation efficiency ratio
can be defined as: ${\cal
M}=\eta_{\textrm{uv}}G_{\textrm{uv}}/G_{\textrm{vis}}$ where $G_{\textrm{uv}}$ and $G_{\textrm{vis}}$ are experimentally
measured excitation functions for one-color illumination (see Fig.S2 in SI), and
$\eta_{\textrm{uv}}$ is the UV pump efficiency $\sim10\%$. The
differential PL spectral function is: $\delta I=\Delta
I/I^{\textrm{vis}+\textrm{uv}}=(1-T^{-1}_{UVLP}
I^{\textrm{vis}}/I^{\textrm{vis}+\textrm{uv}})$ where
$I^{\textrm{vis}}$ and $I^{\textrm{vis}+\textrm{uv}}$ are the
experimentally measured PL intensity under one-color (vis) and two-color
(vis+uv) excitation respectively; $T_{UVLP}$ is the correction
transmission function of the UVLP filter (see Fig.S3 in SI). Using these two measured
functions and known transition rates (Table~S1 in~SI)
 one explicitly obtains $\gamma_{DNA}$,
the additional PL quenching rate appearing under UV illumination of the sample.
Dependence of this non-radiative recombination mechanism,
induced by extra free charge carriers
transferred from the DNA, on the PL and PLE photon energy can be derived
from the experimental 1-color and 2-color QY data as:
\begin{equation}
\gamma_{DNA}= (\gamma_{nr1}+\gamma_{r1})\left(\frac{QY^{1C}}{\tilde{QY}^{2C}}-1\right)
 \label{gDNAlong}
\end{equation}
where the one-color quantum yield is: $QY^{1C}=T^{-1}_{UVLP}
I^{\textrm{vis}}/G_{\textrm{vis}}$ and the effective two-color quantum yield is:
$\tilde{QY}^{2C}=I^{\textrm{vis}+\textrm{uv}}/(G_{\textrm{vis}}+\eta_{\textrm{uv}}G_{\textrm{uv}})$
(reader is referred to SI for detailed derivation).

\begin{figure*}[htb]
	\centering
	{   \includegraphics[width=5.8 in]{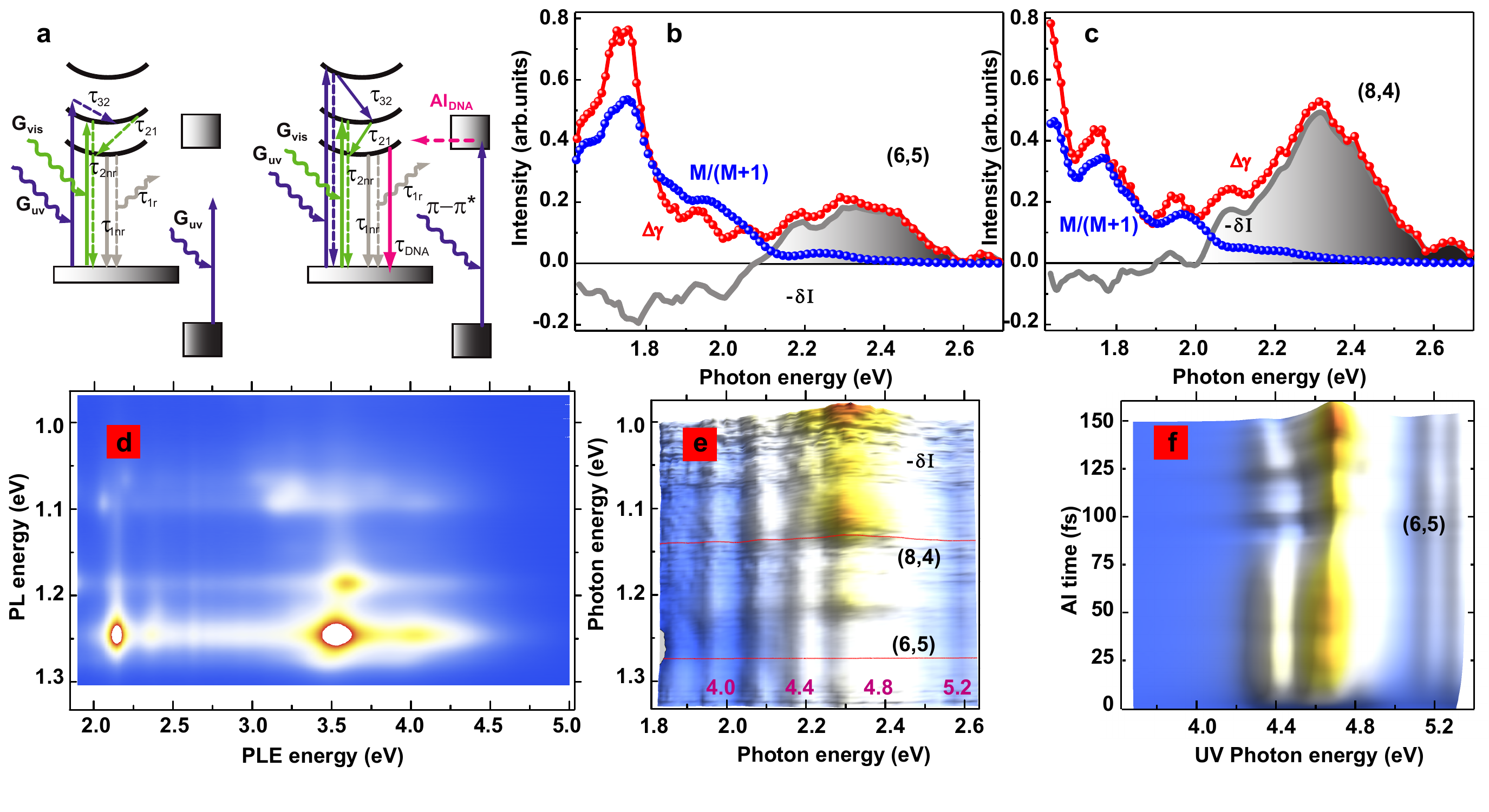}
		\caption{(a) Sketch of electronic transitions after (right) resonant UV excitation of DNA, followed by AI of the $\pi$-$\pi^*$ excited state (red lines); (left) off-resonance excitation, not followed by AI. Blue curves show UV-excitation and corresponding transition rates; green/gray curves correspond to visible/NIR range (see text for details).
			(b-c) Spectral decay function $\Delta_\gamma$ (red), differential PL function $-\delta I$ (gray), and excitation
			ratio function $M(M+1)$ (blue) for (b) the (6,5) SWNT, PL peak at 1.25 eV, bottom red line in panel (e);
			and (c) the (8,4) SWNT, PL peak at 1.1~eV, top red line in panel (e).
			(d) DNA-SWNT PL/PLE map at 1-color excitation.
			(e) Differential PL/PLE map of the same sample as in (d). Cross-sections of $-\delta I$ map at the positions of red lines, corresponding to (8,4) and (6.5) SWNTs, are shown in panels (b-c). Horizontal axis shows double scale: PLE visible range excitation (black) same as in panel (b) above it, and UV pump (purple) same as in panel (f).
			(f) Dynamics of autoionization spectra of the $\pi$-$\pi^*$ excited state of
			DNA-SWNT complex:
			the AI probability of (6,5) SWNT wrapped with (GT)-sequence vs. photon energy of UV pump
			(horizontal axis) and time (vertical). Note the PLE UV photon energy is double the one in the panel (c) above it.
		} \label{fig:fig4} }
\end{figure*}

Fig.\ref{fig:fig4} shows contributions of the individual SWNT species of known
chiralities: panel (b) presents spectral PLE functions detected at $E_{PL}=1.25$~eV
(994~nm), corresponding to (6,5) SWNTs making the largest peak in Fig.\ref{fig:fig2}ab. The
differential PL, $\delta I$, (gray shaded) is negative where the DNA induced
quenching supersedes the UV excitation function (blue). Data collected at
$E_{PL}=1.1$~eV (1130~nm), corresponding to (8,4) SWNTs with much lower concentration
in our sample, is plotted in panel (c) and shows an even clearer signature of the
DNA AI, with the normalized PL decay rate, $\Delta_\gamma = \gamma_{DNA}/(\gamma_{nr1}+\gamma_{r1})$, (red)
reaching $\sim50$\%~at resonance with the
DNA $\pi$-$\pi^*$ transition. Using $\gamma_{nr1}$ and $\gamma_{r1}$ from Table~S1
 in SI
the DNA induced PL quenching rate is estimated to be of the order of 0.01-0.025~ps$^{-1}$ for these 2 SWNTs in
resonance. Other SWNT species demonstrate similar behavior as shown
in Fig.\ref{fig:fig4}e, where the complete PLE map on the DNA induced PL quenching
is presented.

\section{4. Conclusions and Outlook}

Charge transfer from the $\pi$-$\pi^*$ photoexcited states of nucleobases
into the nanotube electronic system has been studied  using a combination of quantum
and statistical theory and two-color PL-PLE characterization. Semiempirical quantum
calculations coupled with a 2-particle Green's function theory were employed to
evaluate the charge transfer rate in an example system: poly(GT) ssDNA wrapped around a (6,5) SWNT. The
calculations indicate a strong coupling between the electronic states of the DNA and
the SWNT and predict a fast (less than 20 fs time scale) autoionization of UV-excited DNA,
dominated by the hole transfer from DNA to the SWNT. The calculation results
indicate that the AI rates are faster than the energy transfer between DNA bases in
oligomers (in the ps-ns range), the charge transfer between the nucleobases in dimers
(in the ps range) and the charge transfer from monomers to the
transient states in the surrounding solvent (in the sub-ps range)\cite{rev-DNA-excit,fast-PI-CT}.

The two-color spectroscopy experiments reveal a strong quenching of the SWNT
photoluminescence when the the UV excitation is resonant with the DNA
$\pi$-$\pi^*$ absorption band, whereas the non-resonant UV excitation leads to a usual
enhancement of the PL quantum yield. The enhancement is resonant with absorption bands of
a particular nanotube, while the quenching is independent of the nanotube chirality. A
rate-equation analysis of the experimental results leads to the PL decay rate on the
order of 40-100 ps, much slower than the calculated
autoionization rate. This suggests that thermalization of the hot carriers transferred from the DNA
is the bottleneck of the autoionization-induced PL quenching.

We propose that two-color (and, potentially, multi-color) pump-probe experiments,
similar to the one presented here, could become a useful technique
for detecting charge transfer in various molecular complexes. In the case of DNA,
decoupling of DNA excitation frequency (UV) from the sensor and pump frequencies
(NIR and visible) enables spectroscopic studies of
DNA ionization under a range of experimental conditions. For example, one could
employ the method for studying autoionization in self-assembled DNA-SWNT complexes,
stable at ambient conditions in water solution and capable of penetrating live
cells\cite{dai-review2015}. Our calculations indicate that DNA-to-SWNT AI is a dominant process
that makes SWNTs ideal reporters in which the problems of previous sensor systems,
for example, due to intra-DNA and
DNA-to-solvent charge transport, have been minimized. Thus, changes in the DNA charged
state could be monitored at room temperature in water solution, eliminating the need
for cryogenic or vapor-phase measurements.

Further developments of the method involve studies of SWNTs of different chirality
wrapped with different DNA. A combination of modeling and experimental studies
similar to ours, could presumably determine the sequence effect on the DNA
autoionization rates and identify the SWNT chirality best suited for AI monitoring from
various DNA sequences. Such a study could reveal the sequence-dependent AI processes
that nature presumably exploits to minimize the UV damage of DNA.

TI and SVR acknowledge support by National Science Foundation (ECCS-1202398 and ECCS-1509786); PS acknowledges REU NSF grant PHY-1359195. AB acknowledges the startup fund support from the University of
Central Florida. The authors gratefully acknowledge access to facilities at the National Institute of Standards and Technology for PL measurements and
the computational time support from the UCF Advanced Research Computing Center STOKES.
We are thankful to Dr. J.~Fagan as the host at NIST, Dr. J.~Reimers for providing us with the CNDO code, and Dr. D.~Roxbury for providing the MD trajectory.


Electronic Supplementary Material: Supplementary
material contains details on the sample preparation and experimental methods, and DNA autoionization modeling; discussion of the alternative recombination mechanisms;
  details on the rate equations and PL quenching rate for DNA-induced non-radiative recombination channel;
the experimentally measured excitation functions for UV-vis pump, and the transmission data for UVLP filter; the rate equation parameters;
  the comparison of the DNA induced PL quenching for 6 different SWNT chiralities. ESM is available in the online version of this article at
  http://dx.doi.org/10.1007/***********************



\newpage
\section*{Supplementary Information: Table of Content}
\begin{enumerate}
	\item Details on DNA autoionization modeling
	\item Discussion of alternative recombination mechanisms
	\item Rate equations and PL quenching rate for DNA-induced non-radiative recombination channel
	\item Experimentally measured excitation functions for UV-vis pump
	\item Transmission data for UVLP filter
	\item Rate equation parameters
	\item Comparison of the DNA induced PL quenching for 6 different SWNT chiralities
\end{enumerate}

\newpage

\section{SI: Details on DNA autoionization modeling}

Although modeling of a DNA-SWNT-water system of the same size as in the experiment is
beyond our current computing abilities, we were able to provide clear theoretical
evidence for fast AI of the $\pi$-$\pi^*$ excited states in DNA wrapped around
a typical SWNT.

\begin{figure}[h]
	\centering
	{   \includegraphics[width=4. in]{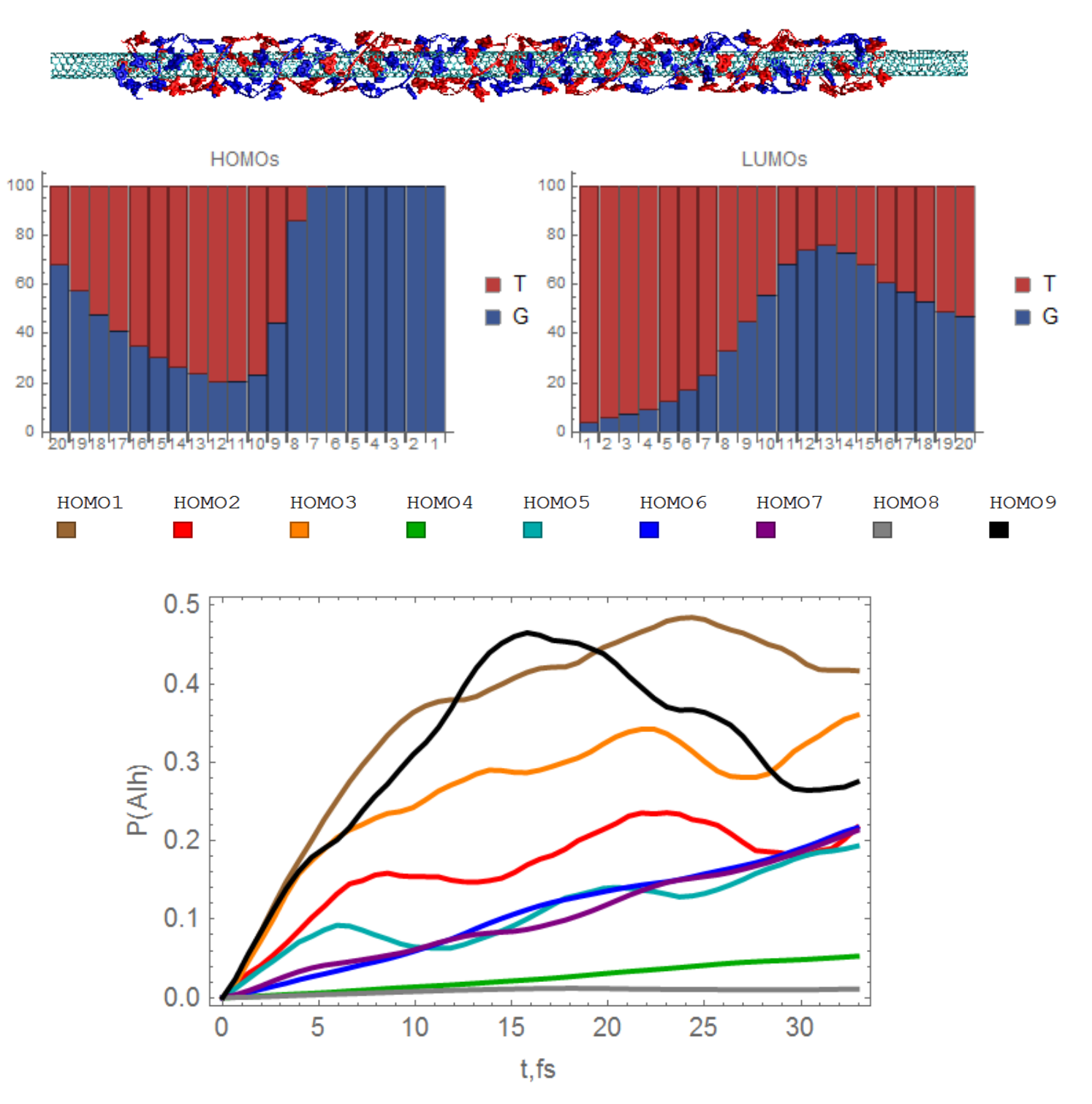}
		\caption{(top inset) Geometry of the whole system. (middle) Statistics of G- and T-base distribution of the smallest energy MOs (20 HOMOs and 20 LUMOs). (bottom) Probability of AI (by hole) for 9 highest occupied MOs for one representative frame of MD.}
		\label{fig:ai-plus-geometry} }
\end{figure}

The most abundant SWNT species in the CoMoCat samples are known to
be (6,5) nanotubes (see Fig.\ref{fig:gf}). The MD simulation was done on a fragment of (6,5) SWNT with
two intertwinned ssDNA oligomers (GT)$_n$ where $n=30$, as shown in Fig.\ref{fig:ai-plus-geometry} (upper inset). Although ssDNA with $n=20$
has been used in our experiments, we note that both olygomers have been shown earlier to make a
similar wrapping on the nanotube surface\cite{jagota}. Frames taken after several ns of MD simulation
were used to determine the geometry of the DNA-SWNT complex solvated in TIP3P
water box. A shorter chain with 14-20
bases (excluding backbone) along with the single unit cell of (6,5) SWNT (364 atoms) were
used for quantum modeling due to computational
expenses. The quantum-chemistry INDO/s calculation has been performed,
followed by a Green's Function based calculation of the autoionization within a
perturbation theory. Time-dependent AI probability for a $\pi$-$\pi^*$ excited DNA base
was calculated (for various configurations) and analyzed (see main text for statistics of AI collected
over the whole trajectory). The AI probability was also convolved with the theoretical
excitation spectrum, giving the total efficiency of UV-ionization, as measured in
the experiment.

Here we provide more details on the quantum-mechanical model. Let us denote by $|b\rangle$ the Dirac states of DNA only, taken in a
water environment, which are solutions of the corresponding Hamiltonian
$H_b$:
\begin{equation}
H_b|b\rangle=E_b|b\rangle
\label{H-DNA}
\end{equation}
where the quantum number $b$ labels non-hybridized (bare) MOs (the energy of the states was renormalized in order to match the experimental absorption band). These
Dirac states are mostly localized on individual bases.

Corresponding states of the isolated SWNT are derived from a similar equation:
\begin{equation}
H_q|q\rangle=E_q|q\rangle
\label{H-NT}
\end{equation}
where the quantum number $q$ includes both linear and angular
momentum, as well as all other required quantum numbers for the SWNT
one-electron energy levels. Those levels are degenerate due to the high
symmetry of the SWNT, except for a few MOs of special symmetry.
The Green's function of the combined system will be defined via the total
Hamiltonian:
\begin{equation}
H|n\rangle=(H_b+H_q+V)|n\rangle=E_n|n\rangle
\label{H-total}
\end{equation}
where $V$ is the interaction/coupling between the DNA and the SWNT, $n$
labels the complete set of hybridized MOs of the complex. We will use below a
special notation to distinguish unoccupied orbitals: $|n^\diamond\rangle$.

This formalism allows us to define the electron autoionization process as an evolution of
the $\pi$-$\pi^*$ excited state into the state with the electron on the SWNT and the
"hole" on the DNA (or vice versa). The initial (many-particle) Dirac state is given by:
$|bb^\diamond,0\rangle$, where we denote the SWNT ground state by zero.
The final state is $\langle b',q_c|$ (correspondingly for hole tunneling
we take $\langle b^\diamond,q_v|$ as the final state). We stress that in both initial and final state
$b,b'$ MOs correspond to the vacant state among normally occupied orbitals
(a hole). Two other states: $b^\diamond$ and $q_c$ belong to normally
occupied (particle) states. The scattering Green's function is as follows (notice that
$\hbar=1$ in this section):
\begin{equation}
G_{b^\prime q_c,bb^\diamond}(t)=\langle b^\prime, q_c|e^{-iHt}|bb^\diamond,0\rangle=\sum_{nn^\diamond, mm^\diamond}\langle b^\prime, q_c|nn^\diamond\rangle
\langle nn^\diamond| e^{-iHt}|mm^\diamond\rangle\langle mm^\diamond|bb^\diamond,0\rangle
\label{green-function-1}
\end{equation}
where in the r.h.s. we use spectral decomposition over the complete
particle-hole basis set. We allow the final state to have
a hole in a different MO $b^\prime\neq b$ (which corresponds to the hole which tunnels to another
base along the DNA chain). Similarly we can write a Green function for the process of an
autoionization of the hole into the SWNT when the electron stays at the DNA in the final state:
\begin{equation}
G^{[h]}_{{b^\diamond}^\prime q_v,bb^\diamond}(t)=\langle {b^\diamond}^\prime, q_v|e^{-iHt}|bb^\diamond,0\rangle=
\sum_{nn^\diamond, mm^\diamond}\langle {b^\diamond}^\prime, q_v|nn^\diamond\rangle
\langle nn^\diamond| e^{-iHt}|mm^\diamond\rangle\langle mm^\diamond|bb^\diamond,0\rangle
\label{green-function-1-hole}
\end{equation}
Neglecting small overlap integrals between occupied and unoccupied states
$\langle m|b^\diamond\rangle, \langle b|m^\diamond\rangle\to 0$, which has been confirmed numerically,
and assuming the thermal equilibrium concentrations of electrons in unoccupied
orbitals: $N^{(n^\diamond)}_e=\langle a^\dagger_{n^\diamond}
a_{n^\diamond}\rangle$ and holes in the occupied orbitals:
$N^{(n)}_h=\langle a_n a^\dagger_n\rangle$ are negligible
compared to unity,
the Green's function Eq.(\ref{green-function-1}) may be written as:
\begin{equation}
G_{b^\prime q_c,bb^\diamond}(t)=\sum_{nn^\diamond}
\langle q_c|n^\diamond\rangle \langle n|b^\prime\rangle\langle b|n\rangle\langle n^\diamond|b^\diamond\rangle e^{-i(E_{n^\diamond}-E_n)t}=G_{q_cb^\diamond}(t)\,G_{bb^\prime}(-t)
\label{green-function-2}
\end{equation}
and it separates into two one-particle Green functions for the electron and
hole components:
\begin{equation}
G_{q_cb^\diamond}(t)=\sum_{n^\diamond}\langle q_c|n^\diamond\rangle \langle
n^\diamond|b^\diamond\rangle e^{-iE_{n^\diamond}t}\qquad
G_{bb^\prime}(-t)=\sum_{n}\langle b|n\rangle\langle n|b^\prime\rangle
e^{+iE_nt}=G_{b^\prime b}^+(t)
\label{single-part-green-function-2}
\end{equation}

The probability of autoionization is:
\begin{eqnarray}
{\cal P}=G(t)G^+(t)={\cal P}_e+{\cal P}_h=G_{b^\prime q_c,bb^\diamond}(t) \, G_{bb^\diamond,b^\prime q_c}(-t)+[holes] \\
=G_{q_cb^\diamond}(t)\,G_{b^\diamond q_c}(-t)\,G_{b b^\prime}(-t)\,G_{b^\prime b}(t)+[holes] \nonumber
\label{prob-AI-01}
\end{eqnarray}

Fig.\ref{fig:ai-plus-geometry} shows the representative time dynamics of the hole AI
for several HOMOs, as labeled in the figure legend. Rabi oscillations are prominent,
overlaid with an overall increase of the probability to find a hole with its final
state on the SWNT (anywhere in the valence band). The middle panel shows statistics for a G- and
T-base population of the 40 MOs with the smallest energy (counted from the charge
neutrality point).

\section{SI: Discussion of alternative recombination mechanisms}

The quantum yield of  PL is proportional to the ratio: $\frac{\gamma_{r1}}{\gamma_{nr1}+\gamma_{r1}}\simeq
\frac{\gamma_{r1}}{\gamma_{nr1}}$, which decreases under
second-color UV illumination. Thus, the non-radiative rate in the lowest
exciton state should increase or the radiative rate should decrease.
According to a detailed equilibrium principle, the latter should be
accompanied with lower absorption (bleaching). UV irradiation was
previously suggested to have a direct influence on the nanotube electronic
structure, for example, via UV-assisted defect creation, SWNT
functionalization or formation of chemical bonds
\cite{doorncrochet11,stranoham09} that lead to lower $\gamma_{r1}$.
This resulted in absorption and PL bleaching, typically long lasting effects
with appreciable recovery times which were not observed in our case. This is
likely because our samples contain tubes densely wrapped with the
(GT) DNA, which is known to cover the SWNT surface densely\footnote{We
	cannot exclude that UV irradiation may influence defects at the ends of the
	tubes which are open to water, however such mechanism should be
	negligible due to the high aspect ratio of our tubes (the length $\sim
	400$~nm).}
without holes, suppressing surface functionalization \cite{DNArecognition-Jagota2012}.

In this work we introduce a new nonradiative mechanism of exciton decay. Here we
briefly discuss several existing mechanisms and argue why they cannot explain our
two-color PL. Three main groups of PL nonradiative decay channels were
considered in the earlier works:

1. At very high excitation intensity, when
the number of excitons per SWNT is larger (much larger) than one,
the exciton-exciton (Auger recombination) decay mechanism \cite{wang006} is
important. A similar exciton-electron Auger recombination was shown
to be less efficient \cite{vasya08,hertelzhu07} than the multi-exciton
channel due to strong restrictions imposed on the symmetry of electronic
states participating in such a transition. These symmetry conditions are
difficult to satisfy in a two-particle process, unless there are bound electrons
in deep (mid-band-gap), strongly localized states or numerous free charge
carriers in conduction or valence bands.

2. For an exciton (or e-h pair) in the
$E_{22}$ (and higher) subbands which overlaps with the single-electron
continuum in a lower subband (for example, $E_{11}$), the
electron-electron \cite{Hertel2010}  (or electron-phonon
\cite{swanHarrah11}) scattering results in an intersubband relaxation. This
non-radiative channel naturally does not apply to $E_{11}$ excitons, lying
below a corresponding single-electron continuum of states.

3. Many-particle
(multi-state) non-radiative channels can become very efficient under the
condition that (a) many-particle interaction satisfies required symmetry
selection rules and (b) the process is resonant in energy, that is, the energy
of the many-particle state coincides with the exciton energy. Such mechanisms
include PAIEI \cite{vasya08} and a number of defect- or impurity-based
exciton scattering models
\cite{swanHarrah11,lefebvreFinnie04,doorncrochet12,harrah11,ghosh10}.

The mechanisms 2 and 3 were not shown to be activated by light.

Because of the protective DNA layer, densely coating the SWNT surface, defect
formation and non-radiative exciton recombination mechanisms based on it
seem unlikely to happen in our experiment. Even though the DNA wrapping could be, arguably,
corrupted during the UV irradiation, we did not detect this effect in our samples.
This would allow access of water molecules to the SWNT surface
thus increasing the non-radiative decay. However, it is inconsistent with the
fact that the anomalous PL was fully reversible and recovered
instantaneously after the second-color illumination was turned off. Also the
solution contained no free DNA, since pure 
heavy water was substituted for the DNA citric buffer in our samples. In the event of DNA
desorption from the tube walls under UV irradiation, its concentration in solution will
be minuscule and  the reverse process of restoration of the wrap will not
happen for entropy reasons. Thus the DNA unwrapping, if it happened, should be easily
detectable in our experiments. We note that we did not observe the effect in
tubes suspended with DOC or SDBS, where the concentration of the free
surfactant in solution is always high and surfactant molecules may indeed
exchange between solvated phase and the phase agglomerated on the tube
surface \cite{doornduque11,roxbury11}.

Excitation power was  $6\times10^{-5}$ light quanta per tube, always
below the limit of creating multiple excitations per tube. Thus any
exciton-exciton interaction mechanisms can be disregarded under main
(visible) excitation, and even more under the weaker UV illumination.
Finally,  the intersubband relaxation mechanisms can be excluded, due to
non-overlap of the $E_{11}$ exciton with its continuum. Then only mechanisms
involving free charge carriers need to be considered.

It is known that additional free charge carriers enhance nonradiative
recombination rates \cite{kinder08,matsuda10}. However, direct
generation of the e-h pairs in the upper SWNT subbands with the UV pump
shows a resonant enhancement at the PLE lines of these subbands and
should vary with the SWNT chirality (not consistent with the data).  Besides
this, the exciton generation still leaves the total charge density of SWNT the
same, and does not break the charge neutrality by itself. Thus, it does not
lead to tube doping and cannot explain the decrease in the PL yield unless
additional ionization and/or trapping mechanisms are invoked.

Therefore, one has
to assume that there is a new mechanism, providing a steady-state doping
of the SWNT, due to a resonant photoexcitation of the tube surrounding.
The photoexcited states should be shallow enough to allow fast
"instantaneous" response upon turning two-color excitation on and off. This excludes
the electrons trapped in solvated states in surrounding water. We
note that the same samples showed an interesting PL behavior during the
formation of the complexes with the multivalent ions\cite{rei-Ignatova} to
be discussed elsewhere.

What could be the origin of these shallow states? We show in the main text that the
resonantly photoexcited DNA undergoes a $\pi$-$\pi^*$ transition, with the
energy of the excitation above the forbidden gap of the DNA $\sim
4.2$~eV, and then transfers the excited charge carrier into the SWNT valence and
conduction bands. Similar charge transfer has been observed earlier in
other SWNT hybrids: with porphyrines \cite{casey08}, fullerenes
\cite{dsouza07}, and organic molecules \cite{hu08}. Indeed the DNA
ionization potential is 8.9-9.3 eV and 8.25-8.62 eV \cite{dna08,dna07}  for
thymine and guanine bases correspondingly, it can vary in different
solvation states. Thus the excited $\pi^*$
states are in resonance with the conduction bands of a typical SWNT,
which has a work function on the order of 5~eV and a band gap
or the order of 1~eV.

We show in the main text that the photoexcited DNA electron can autoionize in the
SWNT bands.
This creates an additional free charge carrier density (although the
whole complex DNA-SWNT is electrically neutral), which facilitates the
electron-electron scattering and non-radiative decay of the $E_{11}$
exciton. Such a mechanism would not vary among SWNTs of different
chirality unless the photoexcited state moves out of the resonant
window, thus decreasing the autoionization rate. This, however,
would require tubes with extremely large band gaps, not found in our
samples.

\section{SI: Rate equations and PL quenching rate for DNA-induced non-radiative recombination channel}

We use rate equations for a 4-level system of exciton manifolds (including the ground state) of the SWNT. Here
$E_{ii}$ where $i=1,2,3$ is the subband index of the lowest exciton subbands, coupled to each other and
to the ground state via the classical rate equations:
\begin{eqnarray}
\begin{array}{l}
\dfrac{dn_1}{dt}=-\gamma_{r1}n_1-\gamma_{nr1}n_1-\gamma_{DNA}n_1+\gamma_{21}n_2
\\ \\ 
\dfrac{dn_2}{dt}=
-\gamma_{r2}n_2-\gamma_{nr2}n_2-\gamma_{21}n_2
+\gamma_{32}n_3+ G_{\textrm{vis}}
\\  \\ 
\dfrac{dn_3}{dt}=
-\gamma_{r3}n_3 -\gamma_{nr3}n_3-\gamma_{32}n_3+ \eta G_{\textrm{uv}},
\end{array}
\label{SI-01}
\end{eqnarray}
where $n_i$ is the exciton population of the $i^{th}$ level (the $E_{ii}$
state), $G_{\textrm{vis}}$, $G_{\textrm{uv}}$ are generating functions
(excitation/pump spectral functions), $\eta$ is the second-color excitation
efficiency, $\gamma_{ij}$ are intersubband relaxation rates, and
$\gamma_{r i}/\gamma_{nr i}$ are the radiative/nonradiative rates of the
corresponding states. Radiative and nonradiative rates were taken from
\cite{Hertel2010,swanHarrah11,schoppler11} and presented in Table
\ref{tab:one}. Generating functions were directly measured with
single-color excitation and presented in Fig.\ref{fig:gf}.

In addition to standard coupling parameters
$\gamma_{ij},\gamma_{ri},\gamma_{nri}$, known from previous works
on monochromatic excitation, we add a new non-radiative decay rate:
$\gamma_{DNA}$
, specific to the nonradiative recombination channel induced  in the $E_{11}$
subband by second-color photoexcitation (UV). We assign this channel
to the photoexcitation of the DNA followed by autoionization in the
$E_{11}$ subband, as described in the main text.

The steady-state left hand side equals zero in Eq.(\ref{SI-01}), allowing a
simple solution for the populations of the states, $n_i$, if the generation
spectral functions $G_{\textrm{vis},\textrm{uv}}$ are known. In
neglecting the radiative transition from all levels except for the lowest one,
the total PL intensity is given by the product of the radiation rate of this
level and it's population factor: $\gamma_{r1} n_1$. After simple math it
can be written as a product of three terms:
\begin{eqnarray}
\begin{array}{c}
I^{\textrm{uv}+\textrm{vis}} =\gamma_{r1}n_i=
\left(G_{\textrm{vis}}+\eta G_{\textrm{uv}}\dfrac{\gamma_{32}}{\gamma_{32}+\gamma_{r3}+\gamma_{nr3}}\right)
\\ \\ 
\times
\left(\dfrac{\gamma_{21}}{\gamma_{21}+\gamma_{r2}+\gamma_{nr2}
}\right)
\left(\dfrac{\gamma_{r1}}{\gamma_{r1}+\gamma_{nr1}+\gamma_{DNA}}\right)
\label{SI-02}
\end{array}
\end{eqnarray}
where the term
$R_3=\gamma_{32}/(\gamma_{32}+\gamma_{r3}+\gamma_{nr3})$
accounts for the relaxation of the charge carriers from the $E_{33}$ subband
into the $E_{22}$ subband, the second term gives a similar internal yield for the
intersubband relaxation of $E_{22}$ charge carriers into the $E_{11}$
subband, and the last term is the PL efficiency in the lowest $E_{11}$ subband.
Here we explicitly included $\gamma_{DNA}$, an additional non-radiative relaxation rate due to
the photoexcited DNA. We assume that the conversion
rate in the upper subbands is large:
$R_3\approx
1/2$ or larger and that it is independent of the excitation energy which
is probably accurate due to the fast and efficient intrasubband
relaxation. Then it can be included together with the UV efficiency in a
single factor $\eta_{\textrm{uv}}$, and we rewrite the solution as:
\begin{eqnarray}
\begin{array}{c}
I^{\textrm{uv}+\textrm{vis}} =  G_{\textrm{vis}}
\left(\dfrac{\gamma_{21}}{\gamma_{21}+\gamma_{r2}+\gamma_{nr2}}\right)
\left(\dfrac{\gamma_{r1}}{\gamma_{r1}+\gamma_{nr1}}\right)
\\ \\
\displaystyle  \times
\left(1+\eta_{\textrm{uv}} \frac{G_{\textrm{uv}}}{G_{\textrm{vis}}}\right)\;
\; \left(\dfrac{\gamma_{r1}+\gamma_{nr1}}{\gamma_{r1}+\gamma_{nr1}+\gamma_{DNA}}\right)
\label{SI-03}
\end{array}
\end{eqnarray}
where the first three terms can be related to the PL intensity at
the single-color (visible) excitation. In order to compare this with the
experimental data one needs to correct the generation spectral function (pump) by
the transmission through the UVLP filter. We had to measure this correction
function ourselves due to a found inconsistency with the nominal transmission
data presented by the manufacturer (see Fig.\ref{fig:uvlp}). Finally, the single-color
PL intensity (while blocking the UV-line with a filter) can be written as:
\begin{equation}
I^{\textrm{vis}} = {\tilde G}_{\textrm{vis}}
\left(\dfrac{\gamma_{21}}{\gamma_{21}+\gamma_{r2}+\gamma_{nr2}}\right)
\left(\dfrac{\gamma_{r1}}{\gamma_{r1}+\gamma_{nr1}}\right)
\label{SI-04}
\end{equation}
where ${\tilde G}_{\textrm{vis}}=G_{\textrm{vis}} T_{UVLP}$ is
the visible excitation spectral function corrected for UVLP filter
transmission.

Two-color PL intensity is then:
\begin{equation}
I^{\textrm{uv}+\textrm{vis}} = \frac{I^{\textrm{vis}}}{T_{UVLP}}
\left(1+\frac{\eta_{\textrm{uv}}G_{\textrm{uv}}}{G_{\textrm{vis}}}\right)\;
\left(\dfrac{\gamma_{r1}+\gamma_{nr1}}{\gamma_{r1}+\gamma_{nr1}+\gamma_{DNA}}\right)
\label{SI-05}
\end{equation}
We did not make any approximation in this expression yet, except for assuming
$\eta_{\textrm{uv}}$ to be independent of energy (which includes an energy
independent factor $R_3$).
It is useful to resolve
Eq.(\ref{SI-05}) for $\gamma_{DNA}$, for which the solution is given by:
\begin{equation}
\gamma_{DNA}=\left(\gamma_{nr1}+\gamma_{r1}\right)\left(
\frac{\frac{I^{\textrm{vis}}}{T_{UVLP}}\left(1+\eta_{\textrm{uv}}\frac{G_{\textrm{uv}}}{G_{\textrm{vis}}}\right)}{I^{\textrm{vis}+\textrm{uv}}}-1\right)
\label{g07}
\end{equation}
We emphasize that all quantities here are the experimentally measured
ones: $I^{\textrm{vis}}$ and $I^{\textrm{vis}+\textrm{uv}}$ are the
PL intensity under one-color (vis) and two-color (vis+uv) excitation
correspondingly; $T_{UVLP}$ is the correction transmission function of the
UVLP filter; $G_{\textrm{uv}}$ and $G_{\textrm{vis}}$ are the excitation
functions for one-color illumination; and $\eta_{\textrm{uv}}$ is the
constant UV pump efficiency $\sim 10\%$.

Total excitation is given by:
\begin{equation}
G_{\textrm{total}}=G_{\textrm{vis}}+\eta_{\textrm{uv}}G_{\textrm{uv}}=G_{\textrm{vis}}\; (1+{\cal M}).
\label{SI-08}
\end{equation}
where we single out the excitation efficiency ratio:
\begin{equation}
{\cal M}=\eta_{\textrm{uv}}\frac{G_{\textrm{uv}}}{G_{\textrm{vis}}}
\label{M-ratio}
\end{equation}
which shows how strong the second-color pump is compared to the main
color source. When ${\cal M}\gg0$ the UV excitation creates a large number
of excitons in higher subbands and gives a considerable rise to the total PL. We also provide, for the
sake of completeness, an expression for the relative efficiency of the dual-color
excitation with respect to the single-color one:
\begin{equation}
\frac{{\cal M}}{1+{\cal M}}=1-\frac{G_{\textrm{1C}}}{G_{\textrm{2C}}}
\label{T-ratio}
\end{equation}

In the main text we defined the differential PL function as:
\begin{equation}
\Delta I=I^{\textrm{vis}+\textrm{uv}}-\tilde{I}^{\textrm{vis}}=I^{\textrm{vis}+\textrm{uv}}-T^{-1}_{UVLP}
I^{\textrm{vis}}.
\label{delta_I-2}
\end{equation}
Corresponding normalized differential PL can be defined as the ratio:
\begin{equation}
\delta I=\frac{\Delta I}{I^{\textrm{vis}+\textrm{uv}}}=\left(1-\frac{T^{-1}_{UVLP}
	I^{\textrm{vis}}}{I^{\textrm{vis}+\textrm{uv}}}\right)
\label{SI-09}
\end{equation}

In the main text we also defined the normalized spectral quenching rate:
\begin{equation}
\Delta_\gamma = \frac{\gamma_{DNA}}{\gamma_{nr1}+\gamma_{r1}}=\frac{\gamma_{DNA}}{\gamma_{1C}^{tot}-\gamma_{DNA}}
\label{SI-a00}
\end{equation}
where $\gamma_{nr1}$ is the non-radiative rate and  $\gamma_{r1}$ is
the radiative recombination rate in the $E_{11}$ subband measured for
one-color (vis) excitation. Then $\gamma_{nr1}+\gamma_{r1}$ and
$\gamma_{1C}^{tot}=\gamma_{DNA}+\gamma_{nr1}+\gamma_{r1}$ are
the total decay rates under one-color and dual-color pumps respectively.

Substituting Eqs.(\ref{SI-04},\ref{SI-05}) into (\ref{SI-09}) and making use
of the short-cut notations given by Eqs.(\ref{M-ratio},\ref{SI-a00}) we
obtain the following formula:
\begin{equation}
\delta I=\frac{1}{1+{\cal M}}({\cal M}-\Delta_\gamma).
\label{delta_I}
\end{equation}
We emphasize that all the terms are non-negative. The anomalous
(negative) differential PL corresponds to ${\cal M}\le\Delta_\gamma$.

Substituting Eqs.(\ref{SI-04},\ref{SI-05}) into (\ref{SI-a00}) and making
use of the same short-cut notations, we obtain the DNA decay rate as a
percent of the one-color total decay rate,
$\gamma_{nr1}+\gamma_{r1}$:
\begin{equation}
\gamma_{DNA}= (\gamma_{nr1}+\gamma_{r1})\left(\frac{T^{-1}_{UVLP}
	I^{\textrm{vis}}}{G_{\textrm{vis}}}\frac{G_{\textrm{vis}}+\eta_{\textrm{uv}}G_{\textrm{uv}}}{I^{\textrm{vis}+\textrm{uv}}}-1\right)
\label{SI-a02}
\end{equation}
which can be simplified for data analysis by using short-cut notations for
one/two-color PL quantum yield:
\begin{equation}
QY_{1C}=\frac{T^{-1}_{UVLP}
	I^{\textrm{vis}}}{G_{\textrm{vis}}} \qquad
QY_{2C}=\frac{I^{\textrm{vis}+\textrm{uv}}}{G_{\textrm{vis}}+\eta_{\textrm{uv}}G_{\textrm{uv}}}
\label{SI-a03}
\end{equation}
which finally results in Eq.(8)
of the main text:
\begin{equation}
\gamma_{DNA}= (\gamma_{nr1}+\gamma_{r1})\left(\frac{QY_{1C}}{QY_{2C}}-1\right).
\label{SI-a04}
\end{equation}

\section{SI: Experimentally measured excitation functions for UV-vis pump}

\begin{figure}[h]
	\centering
	{
		\includegraphics[width=3. in]{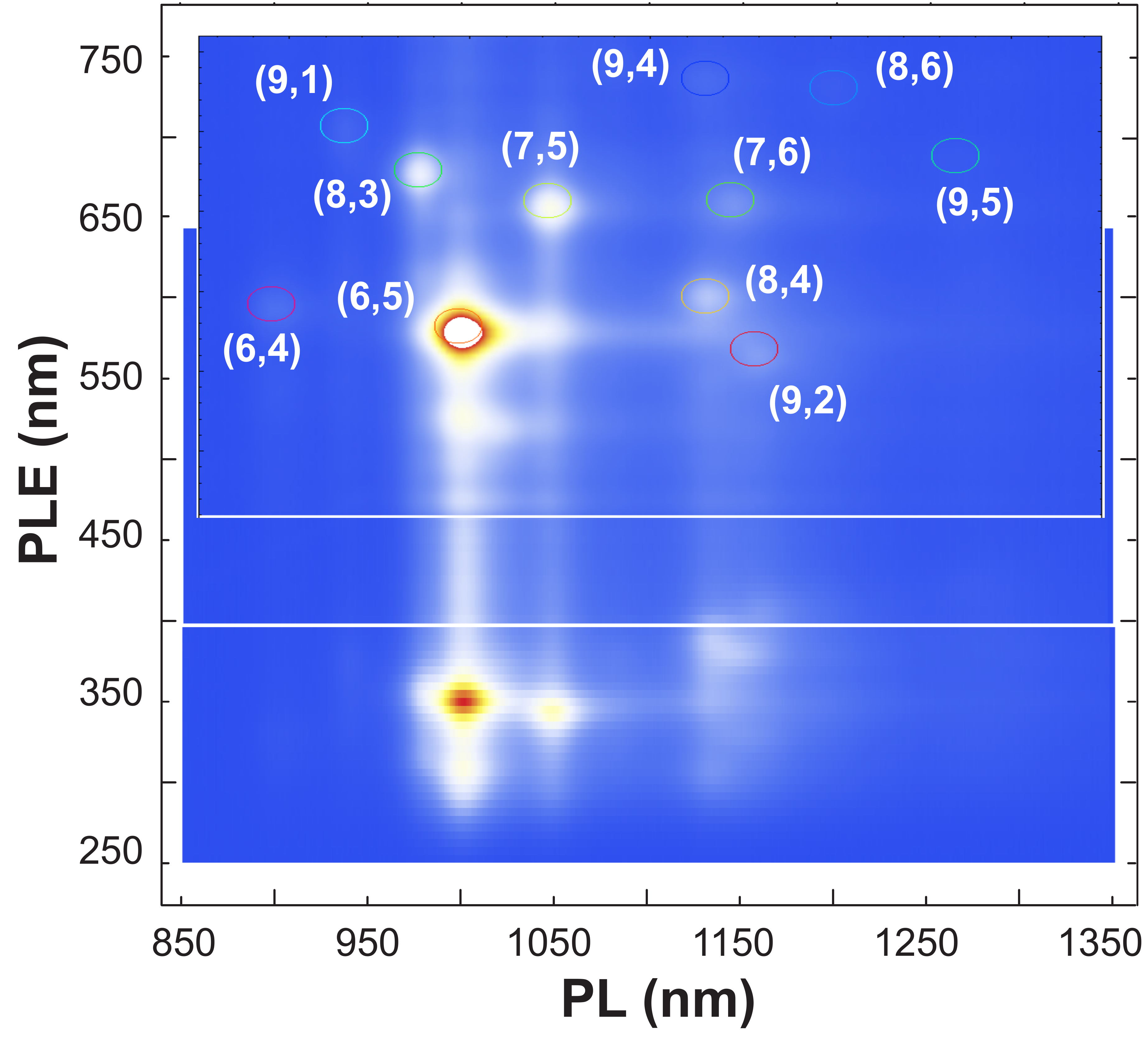}
		\caption{Complete excitation function in the range of PL excitation: 250-760 nm and
			PL range 850-1350 nm is shown.} \label{fig:gf} }
\end{figure}

Experimentally measured excitation functions (generation term in the rate equations) $G^{vis}$ and $G^{uv}$
are presented in Fig.\ref{fig:gf}. The data is combined from 3 separate
measurements, normalized by the total PL excitation strength, following the procedure from\cite{voisin}. Eleven
SWNT species were found in the CoMocat solution: all of the strongest (n,m) peaks in the visible range are
indicated on the Figure.

\section{SI: Transmission data for UVLP filter}

\begin{figure}[h]
	\centering
	{   \includegraphics[width=2. in]{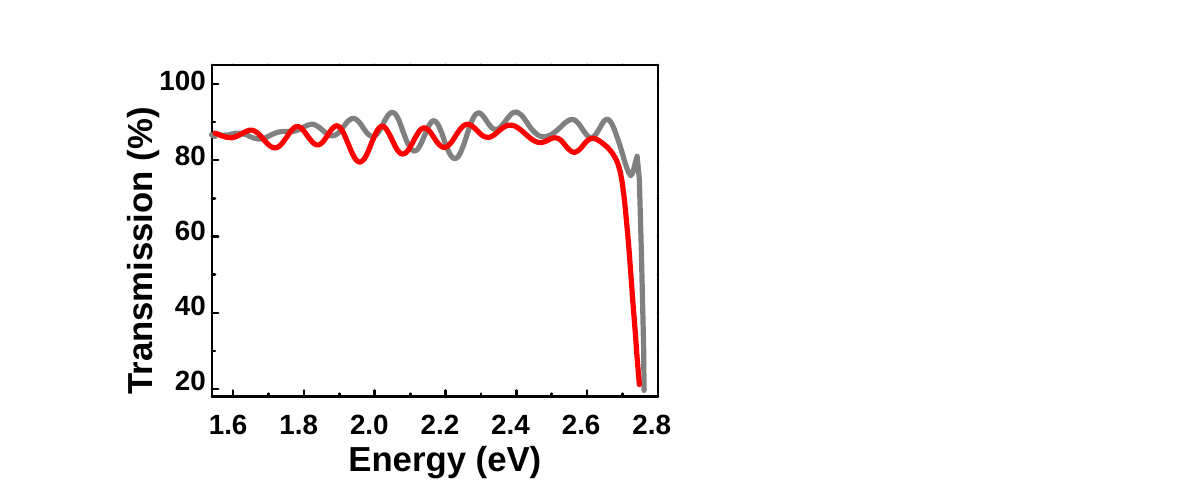}
		\caption{Real UVLP transmission data (red), compared to the nominal transmission, available online (gray).}
		\label{fig:uvlp} }
\end{figure}

Fig.\ref{fig:uvlp} shows the experimental (measured) absorption spectrum of the
UVLP-450 filter, compared to the table data provided by the
manufacturer\cite{thorlabs}. A clear difference is seen between nominal and
actual data which should be taken into account in rate equation analysis.

\section{SI: Rate equation parameters}

All SWNT rates used in the rate equation analysis are given in Table\ref{tab:one} along with
the maximum rates calculated for the DNA induced PL quenching.

\begin{table}
	\caption{Rate equation parameters (all $\gamma$ are in ps$^{-1}$; $~^\star$ for resonant absorption of DNA wrapped on (6,5) SWNT).}
	\begin{tabular}{@{\vrule height 10.5pt depth4pt  width0pt}cccccc}
		\hline
		$\gamma_{r1}$ \qquad & $\gamma_{r2,3}$\qquad &$\gamma_{nr1}$ \qquad &$\gamma_{nr2,3}$\qquad & $\gamma_{21, 32}$ & $\gamma_{DNA}^\star$ \\
		\hline
		6.3$\times10^{-4}$       &10$^{-3}$                 &0.035-0.05            &6.3$\times10^{-3}$           &76.9 &0.017-0.025
		\\
		\hline
	\end{tabular}
	\label{tab:one}
\end{table}

\begin{figure*}[htb]
	\centering
	{   \includegraphics[width=4.5 in]{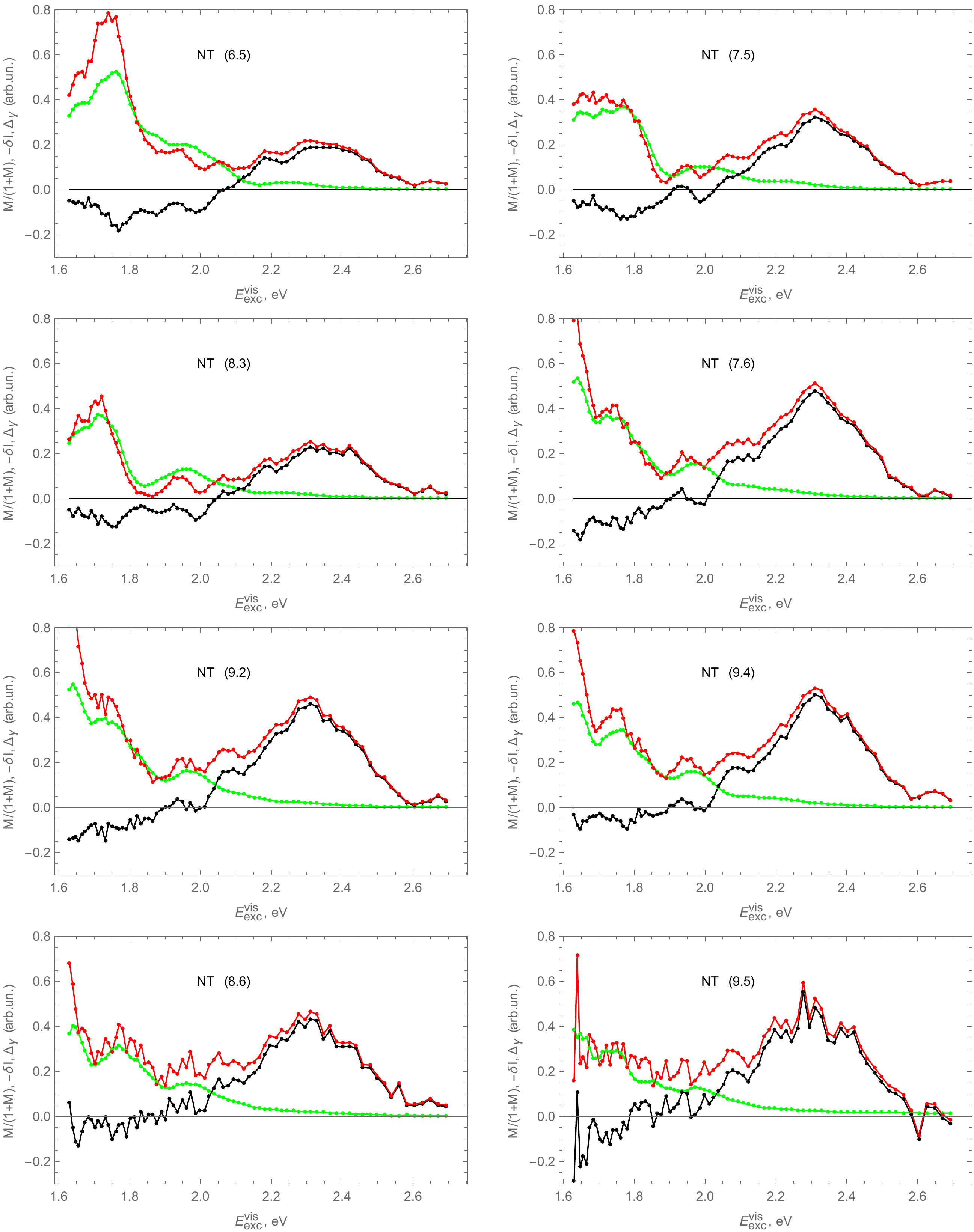}
		\caption{
			The excitation function, $M/(1+M)$, (green),
			the PL difference, $-\delta I$, (black), and
			the PL quenching function, $\Delta_\gamma$, (red).
			Graphics table compares the data for 8 SWNT species from 11 found in solution
			. }
		\label{fig:all-PL-diff} }
\end{figure*}

\section{SI: Comparison of the DNA induced PL quenching for 6 different SWNT chiralities}

The normalized differential
PL, $\delta I$, for a (6,5) nanotube is shown in Fig.\ref{fig:all-PL-diff} along
with the excitation efficiency function ${\cal M}/({\cal M}+1)$ and the normalized function
$\Delta_\gamma$. For a (6,5) tube the DNA induced exciton quenching can
reach above 30\% of the total non-radiative decay rate. Other chiralities
show similar differential PLE spectra with two main features:
on the UV side each $\delta I$ spectrum has a negative band (note that the
abscissa axis is reversed for $\delta I$) around 530~nm where the intensity
of one-color (vis) excitation PL exceeds the two-color PL intensity. This also
corresponds to the region of low efficiency of the UV pump (${\cal M}<<1$).
On the IR side (for ${\cal M}\sim 1$) several features can be found. They
are positioned differently for different tubes following the pattern of
individual resonances of a particular SWNT (compare different chiralities in
Fig.\ref{fig:all-PL-diff}). These features correspond to the resonant UV excitation which
increases the PL signal and compensates for the effect of the DNA (${\cal
	M}>\Delta_\gamma$). The anomalous PL can be observed only for small
${\cal M}$, in the region where the direct UV pump is insufficient compared
to the main excitation line.



\end{document}